\documentclass[journal=jacsat,manuscript=article]{achemso}

\setkeys{acs}{keywords = true} 

\usepackage{chemformula}
\usepackage[mathcal]{euscript}
\usepackage{amsmath}
\usepackage{chemformula}
\usepackage{gensymb}
\usepackage[version=3]{mhchem} 
\usepackage{amssymb}
\usepackage{graphicx}
\usepackage{pdfpages}

\author{Arthur Veyrat}
\author{Valentin Labracherie}
\affiliation{Leibniz Institute for Solid State and Materials Research (IFW Dresden), Helmholtzstraße 20, D-01069 Dresden, Germany}

\author{Dima L. Bashlakov}
\affiliation{B. Verkin Institute for Low Temperature Physics and Engineering, NASU, 47 Nauky Ave., 61103 Kharkiv, Ukraine}

\author{Federico Caglieris}
\affiliation{Leibniz Institute for Solid State and Materials Research (IFW Dresden), Helmholtzstraße 20, D-01069 Dresden, Germany}
\alsoaffiliation{CNR-SPIN, Corso Perrone 24, 16152 Genova, Italy}

\author{Jorge I. Facio}
\affiliation{Leibniz Institute for Solid State and Materials Research (IFW Dresden), Helmholtzstraße 20, D-01069 Dresden, Germany}
\alsoaffiliation{Centro At\'omico Bariloche, Instituto Balseiro and Instituto de Nanociencia y Nanotecnolog\'ia CNEA-CONICET, CNEA, 8400 Bariloche, Argentina}

\author{Grigory Shipunov}

\author{Titouan Charvin}
\affiliation{Leibniz Institute for Solid State and Materials Research (IFW Dresden), Helmholtzstraße 20, D-01069 Dresden, Germany}

\author{Rohith Acharya}
\affiliation{Leibniz Institute for Solid State and Materials Research (IFW Dresden), Helmholtzstraße 20, D-01069 Dresden, Germany}

\author{Yurii Naidyuk}
\affiliation{B. Verkin Institute for Low Temperature Physics and Engineering, NASU, 47 Nauky Ave., 61103 Kharkiv, Ukraine}

\author{Romain Giraud}
\affiliation{Leibniz Institute for Solid State and Materials Research (IFW Dresden), Helmholtzstraße 20, D-01069 Dresden, Germany}
\alsoaffiliation{Université Grenoble Alpes, CNRS, CEA, Grenoble-INP, Spintec, F-38000 Grenoble, France}

\author{Jeroen van den Brink}
\author{Bernd Büchner}
\affiliation{Leibniz Institute for Solid State and Materials Research (IFW Dresden), Helmholtzstraße 20, D-01069 Dresden, Germany}
\alsoaffiliation{Department of Physics, TU Dresden, D-01062 Dresden, Germany}

\author{Christian Hess}
\affiliation{Leibniz Institute for Solid State and Materials Research (IFW Dresden), Helmholtzstraße 20, D-01069 Dresden, Germany}
\alsoaffiliation{Center for Transport and Devices, TU Dresden, D-01069 Dresden, Germany}
\alsoaffiliation{Fakultät für Mathematik und Naturwissenschaften, Bergische Universität Wuppertal, D-42097 Wuppertal, Germany}

\author{Saicharan Aswartham}
\affiliation{Leibniz Institute for Solid State and Materials Research (IFW Dresden), Helmholtzstraße 20, D-01069 Dresden, Germany}

\author{Joseph Dufouleur}
\email{j.dufouleur@ifw-dresden.de}
\affiliation{Leibniz Institute for Solid State and Materials Research (IFW Dresden), Helmholtzstraße 20, D-01069 Dresden, Germany}
\alsoaffiliation{Center for Transport and Devices, TU Dresden, D-01069 Dresden, Germany}

\title{Berezinskii-Kosterlitz-Thouless transition in the type-I Weyl semimetal \ch{PtBi2}}

\SectionNumbersOn

\begin{document}

	\begin{abstract}
		Symmetry breaking in topological matter has become in recent years a key concept in condensed matter physics to unveil novel electronic states.	In this work, we predict that broken inversion symmetry and strong spin-orbit coupling in trigonal \ch{PtBi2} lead to a type-I Weyl semimetal band structure. Transport measurements show an unusually robust low dimensional superconductivity in thin exfoliated flakes up to 126 nm in thickness (with $T_c \sim 275-400$~mK), which constitutes the first report and study of unambiguous superconductivity in a type-I Weyl semimetal. Remarkably, a Berezinskii-Kosterlitz-Thouless transition with $T_\text{BKT} \sim 310$~mK is revealed in up to 60 nm thick flakes, which is nearly an order of magnitude thicker than the rare examples of two-dimensional superconductors exhibiting such a transition. This makes \ch{PtBi2} an ideal platform to study low dimensional and unconventional superconductivity in topological semimetals.
	\end{abstract}

	Van der Waals superconducting materials have attracted a lot of interest in recent years thanks to the development of exfoliation techniques allowing the fabrication of high-quality mono- and few-layer thick structures. Such nanostructures can show a number of interesting properties, such as the emergence of a superconducting state in the two dimensional (2D) limit \cite{Fatemi2018,Cao2018}, a strong increase of the critical temperatures \cite{Li2016c,Rhodes2021} or the enhancement of the critical magnetic field beyond the Pauli limit \cite{Cao2021}. The exact origin and the nature of this superconductivity is likely related to the enhancement of electron-electron interactions at 2D  \cite{Cao2018,Rhodes2021} and it therefore only appears in few-atomic-layer thick films. For such 2D systems, the superconducting state is fundamentally different from the three dimensional (3D) case due to the spontaneous emergence of vortices induced by a Berezinskii-Kosterlitz-Thouless (BKT) transition at reduced dimensions \cite{Berezinskii1972,Kosterlitz1973}. 
	
	Amongst van der Waals superconductors, materials with a non-trivial band structure are intriguing since they may host interesting unconventional superconducting states when the electron spin-degeneracy of Bloch states is lifted, such as mixed singlet and triplet superconductivity \cite{Edelshtein1989,Gorkov2001,Kozii2015} or  Fulde-Ferrell-Larkin-Ovchinnikov (FFLO) finite momentum pairing \cite{Fulde1964,Larkin1965,Mayaffre2014}. In particular, superconducting Weyl semimetals constitute promising systems to study topological superconductivity \cite{Hosur2014,Bednik2015}, and the spontaneous emergence in nanostructures of vortices induced by a BKT transition \cite{Berezinskii1972,Kosterlitz1973} could further allow the investigation of Majorana bound states in the absence of external magnetic field, as recently proposed for iron-based superconductors \cite{Wang2018d,Tang2019}. Despite this great interest, experimental evidences of BKT transitions are scarce and generally limited to some high-quality films due to the sensitivity of the ordered phase to any structural disorder \cite{Brun2014,Brun2016}, and no BKT transition has been observed in any type-I Weyl semimetal so far. The need for high-quality crystal growth and ultra-clean fabrication techniques due to this sensitivity to disorder constitutes one of the main challenges for studying superconductivity at low dimensions.
	
	In this letter, we first predict the presence of type-I Weyl points at 48 meV above the Fermi energy in trigonal \ch{PtBi2} (t-\ch{PtBi2}) and we unveil superconducting properties at ambient pressure and in thin exfoliated flakes. We show that the superconductivity persists at 2D in 41 to 126 nm thin exfoliated samples, and appears more stable when decreasing the sample's thickness. Remarkably, we evidence a BKT transition at unprecedentedly high thickness  in two exfoliated flakes 41~nm and 60~nm thick, making t-\ch{PtBi2} a prime candidate to study the interplay between low dimensional superconductivity and topology, with simple sample fabrication techniques.  Our work constitutes hence the first unequivocal report of superconductivity in a type-I Weyl semimetal, as well as the first experimental evidence of a BKT transition in superconducting Weyl semimetals. 
	
	In the context of exfoliable van der Waals materials, t-\ch{PtBi2} is of particular interest. Beyond the very large linear magnetoresistance measured in the hexagonal \cite{Yang2016} and pyrite \cite{Gao2017} crystal structures, the spin-orbit coupling together with the broken inversion symmetry $\mathcal{I}$ in the layered trigonal structure are responsible for a variety of interesting electronic properties, including a strong Rashba-like spin splitting \cite{Feng2019}, triply degenerate points in the band structure \cite{Gao2018}, signatures of topological edge states at single-layer steps \cite{Nie2020} and pressure induced superconductivity \cite{Wang2021} or under point contact measurements \cite{Bashlakov2022}. 
	
	We calculated the electronic structure of bulk t-\ch{PtBi2} in the space group P31m based on the crystal structure reported in Ref.~\citenum{Shipunov2020}. We performed fully-relativistic density-functional calculations treating the spin-orbit coupling in the four-component formalism \cite{Koepernik1999} (see S1 for more details). Similarly to previous works \cite{Gao2018,Shipunov2020}, the band structure indicates a semimetallic character with several bands crossing the Fermi energy, generating various electron and hole pockets (Fig.\ref{Fig1}.c). A search for accidental crossings of bands at isolated points (Weyl nodes, allowed by broken $\mathcal{I}$) between bands N and N+1, where N is the number of valence electrons per unit cell, yields the existence of twelve symmetry-related Weyl nodes, 48 meV above the Fermi energy (Fig.\ref{Fig1}.b). 
	Fig.~\ref{Fig1} shows the energy dispersion around the Weyl point along the three cartesian directions (Fig.~\ref{Fig1}.d) and the isoenergetic contours around the Weyl point (Fig.~\ref{Fig1}.e). As expected for a type-I Weyl node, the contours reduce to a single point as the energy approaches the Weyl point. Interestingly, the isoenergetic contours of the pocket for energies close to the Weyl point energy are anisotropic in the $(k_\text{y}, k_\text{z})$ plane, with an ellipsoid-like shape which long axis is tilted by a similar angle as the tilt angle corresponding to the maximum of magnetoresistance, possibly indicating a similar origin for both properties.
	\begin{figure}[h!]
		\centering
		\includegraphics[width=0.9\textwidth]{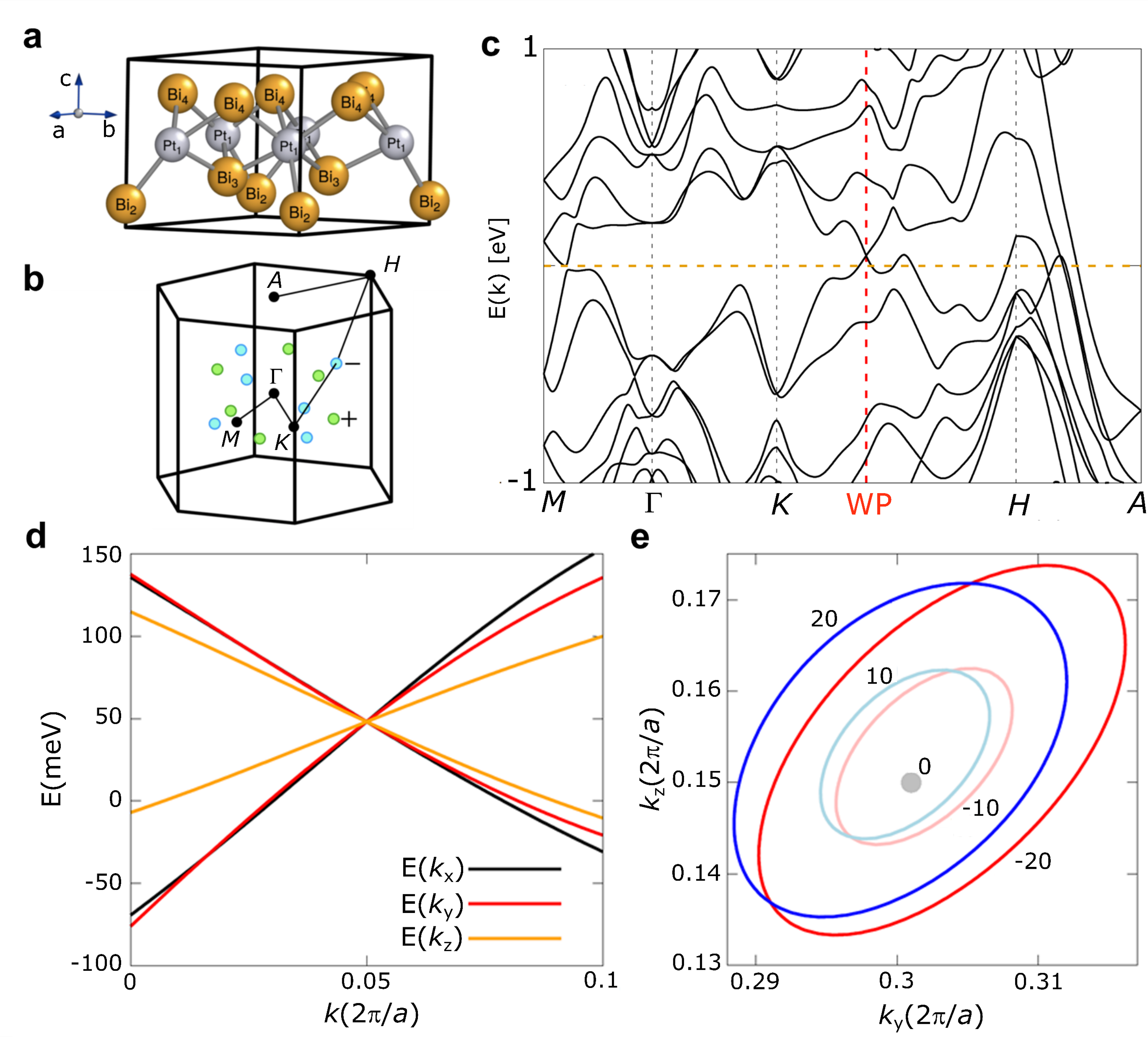}
		\caption{(a) Crystal structure of trigonal \ch{PtBi2}. (b) Brillouin zone. Green (blue) points correspond to Weyl nodes of positive (negative) chirality. (c) Band structure along the path indicated in (b), which includes one of the Weyl nodes located at 48meV above the Fermi energy. (d) Energy as a function of momenta along the three cartesian directions. The path is centered at a Weyl node. It can be seen that the Weyl cone is only weakly canted. (e) Isoenergetic contours fixing the energy 0, $\pm 10$~meV and $\pm 20$~meV with respect to the Weyl point energy.}
		\label{Fig1}
	\end{figure}

	To investigate the properties of t-\ch{PtBi2}, high quality single crystals were grown via the self-flux method \cite{Shipunov2020} and contacted in a four-probe configuration using silver wires and conducting epoxy. The temperature and magnetic field dependence of the resistance, as well as the measured Shubnikov-de Haas oscillations, show a good agreement with previously reported results \cite{Gao2018} as well as with our band structure calculations (S2).
	
	Thin t-\ch{PtBi2} flakes of around $10~\mu$m wide and few tens~nm thick were exfoliated and contacted with standard e-Beam lithography techniques. We performed a small Ar-etch to remove any surface-oxidation before taking the contacts. This surface oxidation is not expected to influence the transport properties, as it should be dominated by bulk states. Four devices D1, D2, D3 and D4 with respective thickness 60~nm, 126~nm, 70~nm and 41~nm (see Fig.S9 and Fig.S13 for optical pictures of the samples, as well as current and magnetic field orientations) were measured down to $100$~mK in a $^3$He-$^4$He dilution refrigerator (S3). The residual resistance ratio (RRR), which measures the ratio between the resistivity at 300~K and 4~K, indicates that the disorder strength increases continuously with the reduction in flake thickness. Below 1~K, all four nanostructures exhibit a superconducting transition with critical temperatures $300$~mK$\leq T_c \leq 400$~mK and  critical currents $I_c \sim 10-20 ~\mu$A at $100$~mK. The critical temperature is defined here by $R(T_c) = R_N/2$, with $R_N$ the resistance in the normal regime, the critical current $I_c$ and fields $B_c$ being defined similarly  (S4, S5). 
	
	\begin{figure}[t]
		\centering
		\includegraphics[width=0.45\textwidth]{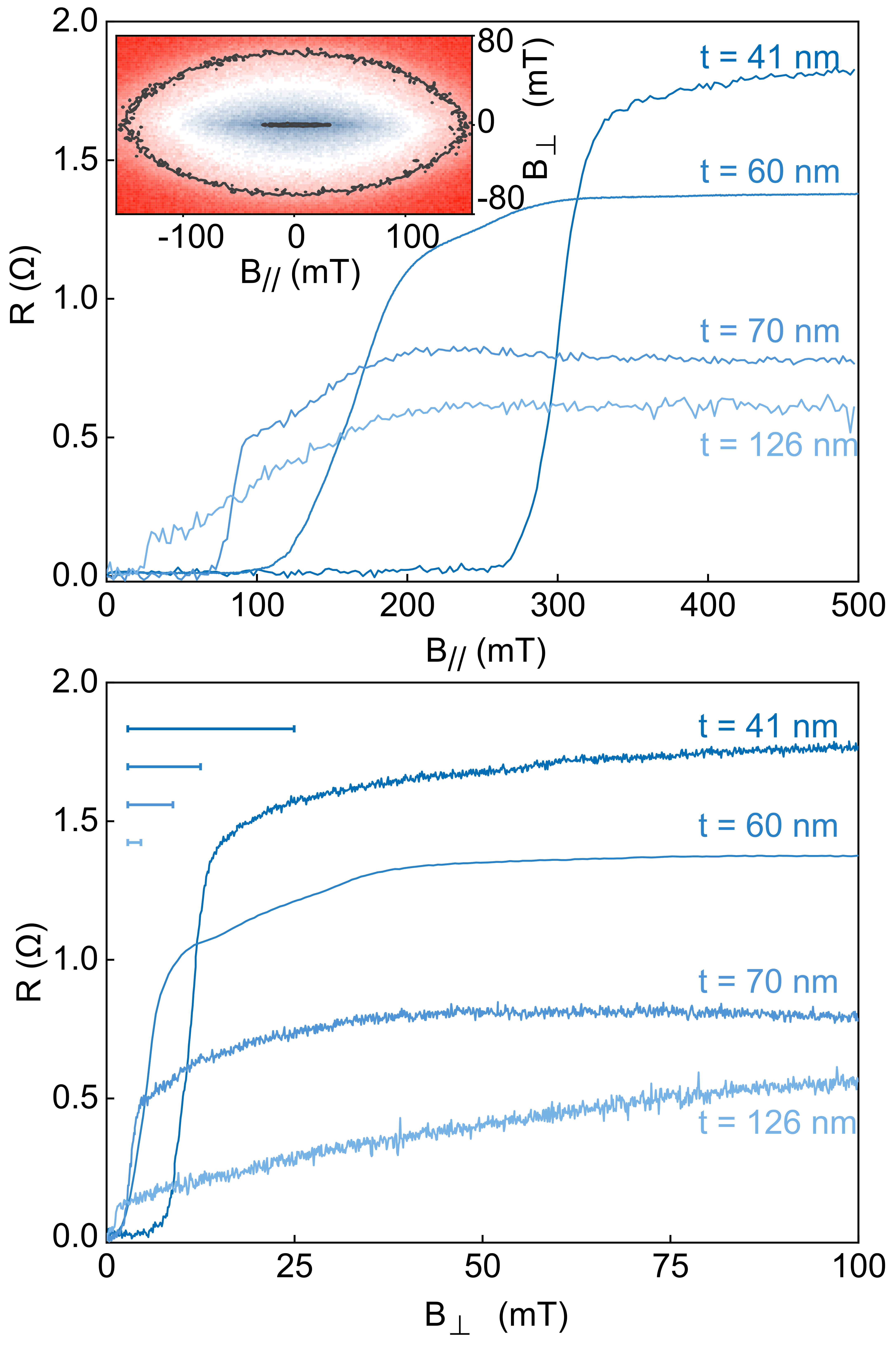}
		\caption{Magnetoresistance of D1, D2, D3 and D4 with the magnetic field applied in-plane ($B_\parallel$, upper panel) and out-of-plane ($B_\perp$, lower panel) measured at $T\sim 100$mK. The inset shows a mapping along the in-plane and out-of-plane magnetic field for D2 (126~nm) with the two line levels corresponding to the slightly anisotropic broad transition and the strongly anisotropic abrupt transition. The perpendicular magnetorestistance are shifted by $3.25$~mT in order to account for the remanent field of the coil. The bar lines indicate the values of $2B_{c,\perp}$.}
		\label{Fig2}
	\end{figure}
	
	We measure a strong dependence on the sample's thickness for both the out-of-plane and the in-plane magnetic field responses of t-\ch{PtBi2}: while the thinnest sample (41nm, D4) shows a single abrupt transition to the normal resistance $R_N$, thicker nanostructures undergo a first abrupt transition to an intermediate resistance $R_0 < R_N$ at a low magnetic field (which decreases with increasing thickness), followed by a second broad transition to $R_N$ at higher field (Fig.~\ref{Fig2}). This broad transition is reminiscent of a macroscopic single-crystal which undergoes a broad superconducting transition around $T_c \sim 600$~mK (S6). The broad transition is found to be almost isotropic as in a macroscopic crystal (S6) whereas the abrupt transition is strongly anisotropic, as shown in the inset of Fig.\ref{Fig2}. The slight anisotropy of the broad transition which can still be seen in the inset can be attributed to misalignment issues of the sample in the magnetic fields, as the transition becomes isotropic when mapped along the perpendicular in-plane direction. We concentrate below on sample D1 (60nm thick), the results obtained on the three other samples (D2, D3 and D4) being shown in S5 and S7.
	
	\begin{figure}[t]
		\centering
		\includegraphics[width=0.45\textwidth]{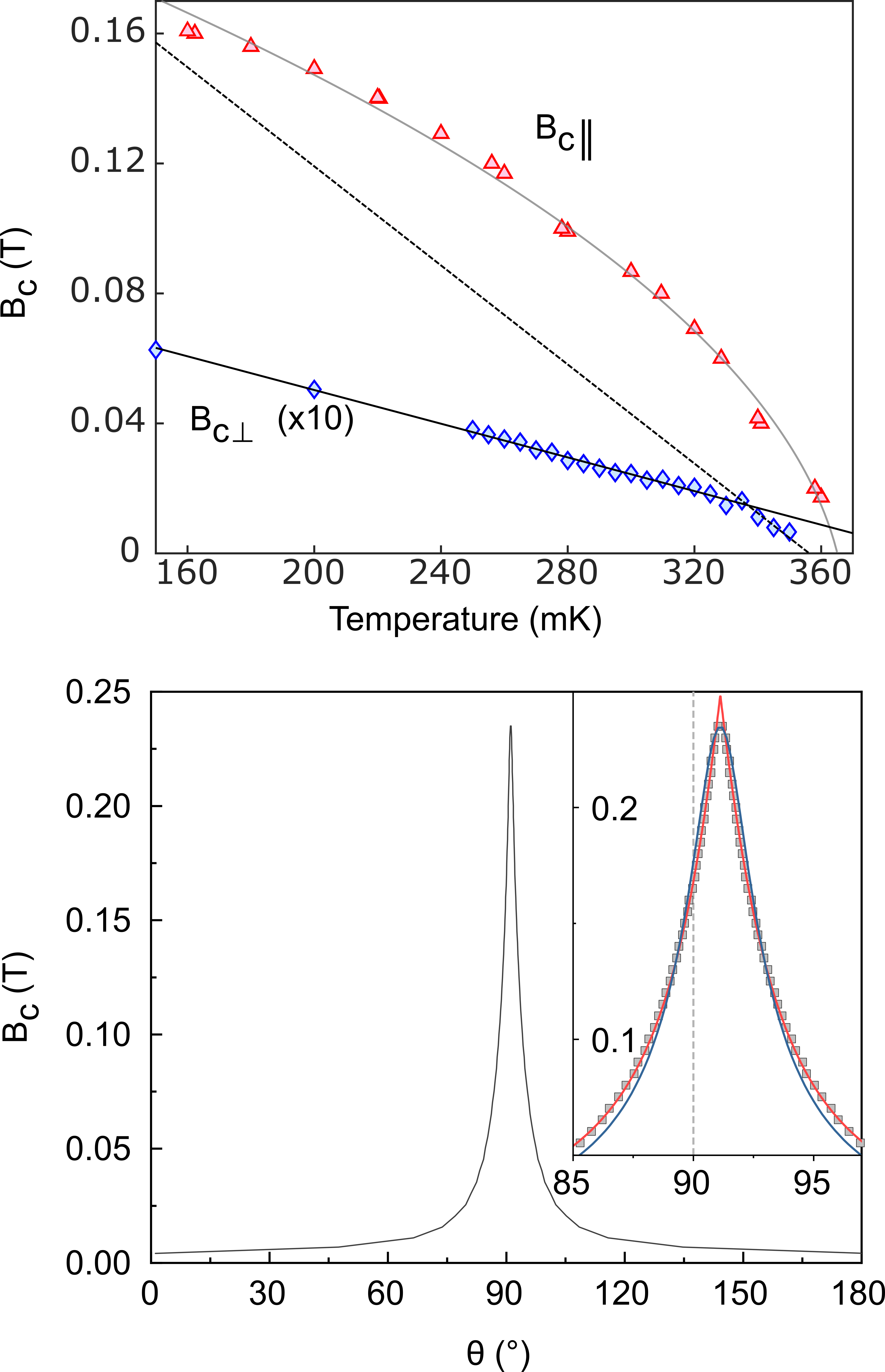}
		\caption{Upper panel: Temperature dependence of the out-of-plane (blue diamonds) and in-plane (red triangles) critical fields, fitted with the 2D Ginzburg Landau model, including the misalignment (see S9). The out-of-plane critical field is increased tenfold for visibility and the fit at low and high temperature are indicated with the plain and dashed line respectively. Lower panel: Angular dependence of $B_c$ at T=100 mK. Inset: Experimental data points (grey squares) and fits with the 3D GL model (blue line) and the 2D Tinkham model (red line).}
		\label{Fig3}
	\end{figure}
	
	The out-of-plane critical magnetic field (along the c-axis) $B_{c,\perp}$ in D1 is about 6~mT at 100mK. It depends linearly on the temperature over a wide temperature range below $T_c$ (Figure \ref{Fig3} up, in blue), as expected from the Ginzburg-Landau theory \cite{Tinkham1975}:
	\begin{equation}
		\label{eq:BcT_3D}
		B_{c,\perp} (T) =\frac{\phi_0}{2\pi \xi_{||}^2}(1-T/T_c ),
	\end{equation}
	with $\phi_0$ the superconducting flux quantum and $\xi_{||}$ the average in-plane superconducting phase coherence length at $T=0$. Fitting the data with Eq.~\ref{eq:BcT_3D} yields $\xi_{||}=180$~nm and $T_c=392$~mK. At 335mK, $B_{c,\perp}(T)$ changes its slope abruptly, corresponding to $\xi_{||}=120$~nm and $T_c=360$~mK. A clear explanation for this unexpected crossover is still lacking so far, although it might be related to inhomogeneities in the sample. We note that the latter value of $T_c$ is more consistent with other measurements performed on this device.
	
	The critical field is strongly anisotropic when tilting the field from the out-of-plane direction ($B_{c,\perp}$) to the in-plane direction ($B_{c,\parallel}$) with a ratio $B_{c,\parallel} / B_{c,\perp}$ as large as 57 at $T=100$mK in D1, considering a $1.1\degree$ misalignment of the sample (see Fig.~\ref{Fig3} for D1 and S7 for D2, D3, and D4). $B_{c,\parallel}(100\text{mK}) \sim 240$mT remains below the Pauli paramagnetic limit \cite{Clogston1962,Chandrasekhar1962} given by $B_p \sim 1.84 T_c \sim 660$mT. Considering the sample's misalignment (see Eq. 1 in S9), the temperature dependence is well fitted by the 2D phenomenological Ginzburg-Landau theory (Figure \ref{Fig3} upper panel, in red, as well as S8 and S9):
	\begin{equation}
		B_{c,\parallel}(T)=\frac{\phi_0 \sqrt{12}}{2\pi\xi_{\parallel,1} t_\mathrm{SC}}\sqrt{1-T/T_c},
		\label{Bc_para}
	\end{equation}
	with $t_\mathrm{SC}$ the thickness of the superconducting layer and $\xi_{\parallel,1}$ the in-plane coherence length in the direction perpendicular to $B_{\parallel}$. We cannot however fit $B_{\parallel}(T)$ satisfyingly only by injecting the value of $\xi_{\parallel} = 120$nm extracted from the $B_{\perp}(T)$ measurements, due to in-plane anisotropies in the superconducting state (as already measured in \ch{MoTe2} \cite{Cui2019}). We therefore consider two in-plane phase coherence lengths $\xi_{\parallel,1} \neq \xi_{\parallel,2}$ in orthogonal in-plane directions, with $\xi_{\parallel} = \xi_{\parallel,1}*\xi_{\parallel,1}$ in Eq.~\ref{eq:BcT_3D}, and fit the data from $B_{c,\perp}$ and $B_{c,\parallel}$ together. When fixing $t_\mathrm{SC} = 55$nm, to account for potential surface oxidation of the sample over a couple nanometres on each side (atomic force microscopy measurement give a thickness of $t_{\text{AFM}}=60$~nm), we obtain an excellent fit yielding $\xi_{\parallel,1}=83$~nm and $\xi_{\parallel,2}=146$~nm, with a small anisotropy of $\xi_{\parallel,2}/\xi_{\parallel,1} \sim 1.8$. We note that choosing $t_{SC} \neq 55$~nm increases this anisotropy. We also obtain very similar critical temperatures between in- and out-of-plane directions, respectively $T_{c,\perp} \sim 356$~mK and $T_{c,\parallel} \sim 365$~mK. The small discrepancy between the two temperatures can be explain by some small hysteretic behaviour in the transition which go beyond the scope of this paper and will be treated separately in a future publication.
	This indicates that superconductivity can be attributed to confined bulk states (i.e. $ \xi_{\perp} > t_{\text{AFM}}$) rather than to surface states, constituting a manifestation of the 2D nature of the superconductivity in nanostructures.
	Although Eq. \ref{Bc_para} is only valid for $T \lesssim T_\text{c}$, it fits very well the experimental data over the full temperature range in Fig.~\ref{Fig3} for device D1, in contrast with devices D2, D3, and D4, which show a good agreement only close to the superconducting transition, as discussed in S9.
	
	The angular dependence $B_{c}(\theta)$ ($\theta=0\degree$ corresponding to an out-of-plane field) provides further evidence of the reduced dimensionality of the superconductivity. At 2D, the Tinkham model predicts a cusp-like peak in $B_c(\theta)$ at $\theta= 90\degree$ (Eq.~\ref{Tinkham_2D}), contrarily to the 3D anisotropic Ginzburg-Landau model (Eq.~\ref{GL_3D}) \cite{Tinkham1975}:
	\begin{align}
		\left\lvert \frac{B_c(\theta)cos(\theta-\theta_0 )}{B_{c,\perp}} \right\lvert + \left( \frac{B_c(\theta)sin(\theta-\theta_0 )}{B_{c,\parallel}} \right)^2 &=1 \label{Tinkham_2D}  \\
		\left( \frac{B_c(\theta)cos(\theta-\theta_0 )}{B_{c,\perp}} \right)^2 + \left( \frac{B_c(\theta)sin(\theta-\theta_0 )}{B_{c,\parallel}} \right)^2 &=1; \label{GL_3D}
	\end{align}
	where $\theta_0$ accounts for the misalignment angle of the sample with respect to the magnetic field plane, so that $B_{c,\perp}=B_c(\theta_0)$ and $B_{c,\parallel}=B_c(\theta_0 + 90\degree)$. Our measurements show a very sharp peak around $\theta \sim 91.1\degree$ (see Figure \ref{Fig3}, low), from which we determine $\theta_0 \sim 1.1\degree$. Because of the strong anisotropy measured in the sample, this results in a 25\% reduction of $B_c$  at $\theta=90\degree$ with respect to the true in-plane critical field. The zoomed-in inset shows the comparison between fits for the 2D (red) and 3D (blue) models, giving a better fit for the 2D model. Similarly, the 2D nature of the superconductivity was evidenced in all three samples D2, D3 and D4 (S7). Remarkably, these measurements confirm the 2D nature of the superconductivity even for thicknesses as large as 126~nm in D2, for which $\xi_0 \sim t$.
	
	\begin{figure}[t]
		\centering
		\includegraphics[width=0.47\textwidth]{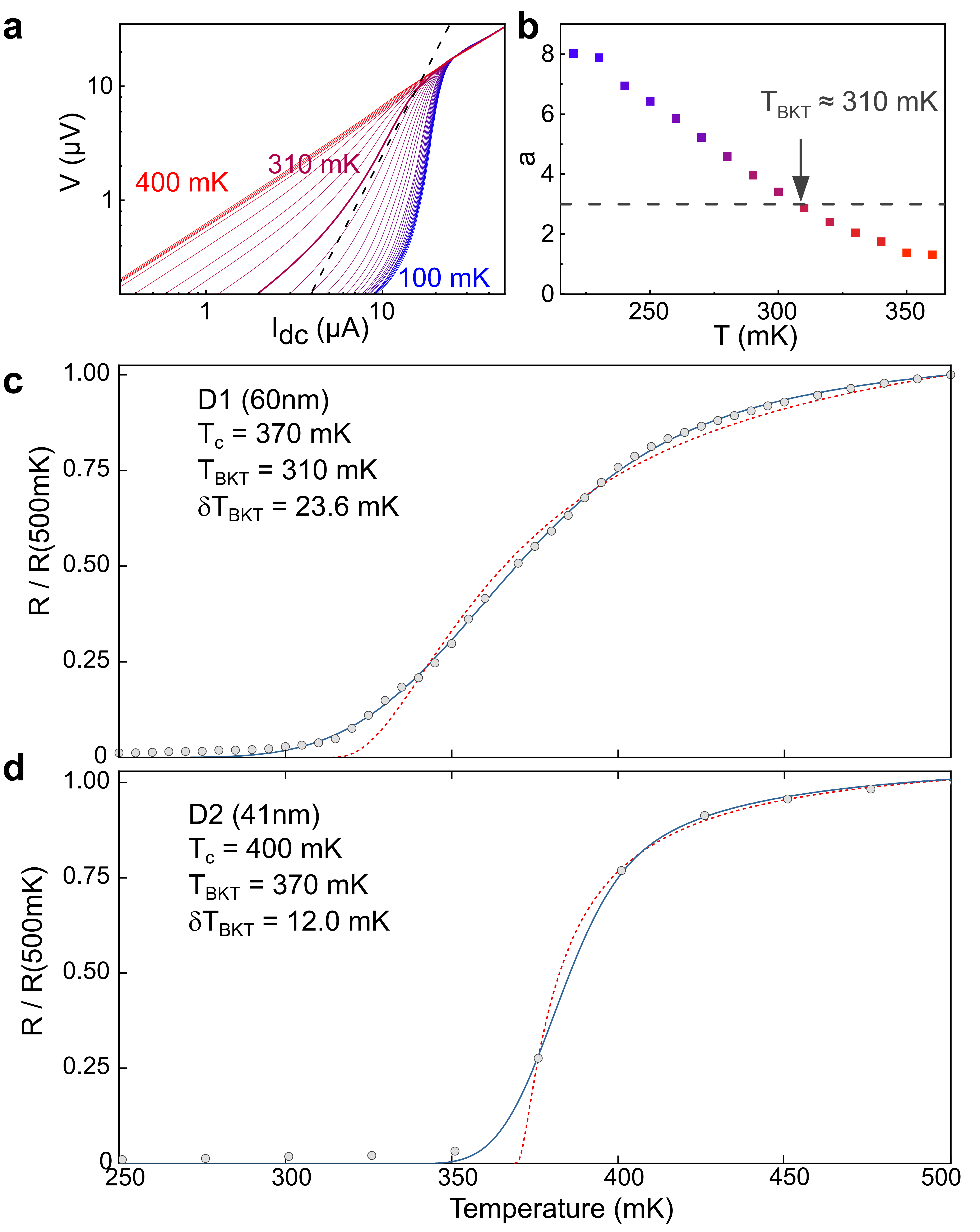}
		\caption{(a) $V(I)$ characteristics for different temperatures, in logarithmic scale and for D1. The dashed line corresponds to $V \propto I^3$. (b) Temperature dependence of the exponent $a(T)=1+\pi J_S(T)/T$ in D1. The value $a=3$ is indicated by a dashed line. (c) Temperature dependence of the resistance (grey circles), fitted with the Halperin-Nelson theory (best fit in red dashed line)
			and the Benfatto model (blue line, where $T_c=370$mK and $T_\mathrm{BKT} = 310$mK are fixed parameters
			and giving $\delta = 24$mK). (d) Temperature dependence of the resistance in D4 (grey circles), fitted with the Halperin-Nelson theory (best fit in red dashed line) and the Benfatto model (blue line, giving $\delta = 12$mK).}
		\label{Fig4}
	\end{figure}
	
	Due to the low dimensionality of the superconductivity in nanostructures, we further sought signatures of a BKT phase transition, as revealed by $R(T)$ and $V(I)$ measurements. Importantly, the measurement of such a transition allows to quantify the degree of inhomogeneities of the system. As shown in Figure \ref{Fig4}, non-linearities appear at low temperature in the $V(I)$ characteristics, with a temperature-dependent power law \cite{Benfatto2009}: $V \propto I^{a(T)}$, with $a(T)=1+\pi J_S(T)/T$ and $J_S$ the superfluid density. The exponent $a$ is equal to $1$ in the normal state ($T \geq T_c$  or $I \gg I_c$), corresponding to Ohm's law. In the case of a BKT transition in a homogeneous and infinite sample, a universal jump of $a$ is expected at the BKT temperature, from $a(T_\mathrm{BKT}^+)=1$ to $a(T_\mathrm{BKT}^-)=3$, and $a(T)$ continuously increases with decreasing temperature for  $T<T_\mathrm{BKT}$. In practice however, for inhomogeneous and finite samples, the discontinuity in $a(T)$ at $T_\mathrm{BKT}$ is smoothed out between $T= T_\mathrm{c}$ and $T=T_\mathrm{BKT}$ and no universal jump is expected \cite{Benfatto2009}. Hence, $a(T)$ slowly increases when T decreases below $T_{c}$ and largely exceeds 3 at temperatures much below $T_\mathrm{BKT}$. The BKT transition occurs when $\pi J_S(T_\mathrm{BKT})/T_\mathrm{BKT} =2$ so that $a(T_\mathrm{BKT})=3$, a relation which defines $T_\mathrm{BKT}$.
	
	We observe such a behaviour in the $V(I)$ characteristics as shown in Fig.\ref{Fig4}a,b where a cubic power law is measured for $T_\mathrm{BKT} \sim 310$mK. We note that due to the low resistance of our sample, the voltage remains significantly smaller than the broadening of the Fermi-Dirac distribution ($\sim 4k_\mathrm{B}T$) for the DC currents applied ($eV_{dc}<20 \mu$eV) so that measurements are done very close to equilibrium. This rules out heating issues due to electron-electron interactions.
	
	To further confirm that a BKT transition occurs in our sample, we investigated the temperature dependence of the resistance. Following the Halperin and Nelson's theory \cite{Halperin1979}, Benfatto et al.  \cite{Benfatto2009} derived a temperature dependence of the resistance for $T \geq T_\mathrm{BKT}$  given by $R(T)/ R_\mathrm{N} = 1/[1+\Delta \sigma / \sigma_\mathrm{N}]$, with 
	\begin{equation*}
		\Delta\sigma / \sigma_\mathrm{N}=4 / A^2 \times \left[\sinh\left( 2 \alpha \sqrt{T_c - T_\mathrm{BKT}}/\sqrt{T - T_\mathrm{BKT}}\right)\right]^2
	\end{equation*}
	where $\sigma_\mathrm{N} = 1/R_\mathrm{N}$, $A$ is a constant of the order of unity and $\alpha$ is the scale of the vortex-core energy, which may deviate from the value expected in the XY model ($\alpha =1$). For $T < T_\mathrm{BKT}$, we have in this case $R(T)=0$, however no reasonable fit could be obtained when fixing $T_\mathrm{BKT}$ to the value given by the $V(I)$ characteristics, $T_\mathrm{BKT} = 310$mK (Fig.\ref{Fig4}.a,b). Benfatto et al. proposed an extended model accounting for inhomogeneities by introducing a Gaussian distribution of the critical current, or equivalently of $T_\mathrm{BKT}$:
	\begin{equation}
		\dfrac{R(T)}{R_\mathrm{N}} = \dfrac{1}{\sqrt{2 \pi} \delta} \int r(t) \text{exp} \left( - \dfrac{(t-T_\mathrm{BKT})^2}{2 \delta^2} \right) \mathrm{d}t
		\label{Benfatto}
	\end{equation}
	with $r(t)=(1 + 4/A^2 \left[ \text{sinh} \left( b \sqrt{t} / \sqrt{T-t} \right) \right]^2 )^{-1}$ and ${b \sim 2\alpha \sqrt{(T_c -T_\mathrm{BKT})/T_\mathrm{BKT}}}$. The experimental data  in Fig.\ref{Fig4}.c can be very well fitted by Eq. \ref{Benfatto}, even for $T_c$ fixed at 370mK, as determined above, and taking $T_\mathrm{BKT} = 310$mK from the $V(I)$ characteristics. It yields a minimal Gaussian spread of $T_\mathrm{BKT}$ of $\delta \sim 24$mK, which compares very well with the spatial distribution of $T_\mathrm{BKT}$ measured between different sets of contact pairs ($T_\mathrm{BKT}=310-340$~mK in S10). To our knowledge, and excluding the case of effective quasi-2D layered superconductors \cite{Baity2016,Guo2017}, our flake is five times thicker than any superconducting films reported so far exhibiting such a transition  \cite{Mondal2011}. 
	Through a similar analysis, we also confirm the presence of a BKT transition in the 41~nm thick sample D4 at $T_\mathrm{BKT} \sim 370$mK (see Fig.\ref{Fig4}d), with a lower Gaussian spread of transition temperatures of ${\delta \sim 12\text{mK} \pm 7}$mK. The error bar in the determination of $\delta$ being of the same order of magnitude as $\delta$ itself, this points to a very weak effect of the broadening. This is confirmed by the very good fit of the data with the Halperin-Nelson formula that stands for the homogeneous case with no broadening, and indicates that the superconducting state in D4 is close to the pure superconducting state. Remarkably, this weak impact of inhomogeneities on the BKT transition broadening is in stark contrast with the lower RRR and $T_c$ in D4 compared to D1, which indicates a stronger disorder in the former (see Table S1). This suggests that both the disorder and the thickness might play key roles in stabilizing the superconducting state at low dimension in t-\ch{PtBi2}.
	
	The discovery of type-I Weyl nodes in the band structure by DFT calculations and the measurement of a superconducting state shed some new light on t-\ch{PtBi2}, which constitutes therefore the first type-I Weyl semimetal with unambiguous superconducting properties. Moreover, we evidence for the first time BKT transitions in thin exfoliated flakes of Weyl semimetals. The persistence of the BKT transition in unusually thick flakes, up to 60~nm, is a remarkable property of t-\ch{PtBi2}, our thickest flake exhibiting a BKT transition at 5 times the thickness of the thickest 2D superconductor with such a transition reported so far \cite{Mondal2011}. Beyond the relevance of t-\ch{PtBi2} to study unconventional superconductivity, we finally stress that a BKT transition in t-\ch{PtBi2} is associated with the emergence of pairs of vortices and anti-vortices at zero magnetic field which have been predicted to host Majorana bound states \cite{Wang2018d}. 


	\newpage
	\begin{titlepage}
		\centering{
			\LARGE{
				\textbf{Berezinskii-Kosterlitz-Thouless transition in the type-I Weyl semimetal \ch{PtBi2}}
				\vspace{10mm}
				\\ Supporting Information}}
		\tableofcontents
	\end{titlepage}
	
	\newpage
	
	\renewcommand{\thesection}{S\arabic{section}}
	\renewcommand\thefigure{
		S\arabic{figure}}

	\section{\label{BandStructure} Band structure calculations}
	
	\textbf{DFT calculation.} We performed DFT calculations using the FPLO code \cite{Koepernik1999}, based on the Perdew Burke Ernzehof generalized gradient approximation. Brillouin zone integrations were done based on a linear tetrahedron method together with k-mesh of 12x12x12 subdivisions. The calculation of the Weyl nodes and of the surface Fermi surface were performed based on a tight-binding model obtained by constructing Wannier functions with the projective technique implemented in the FPLO code. The model includes the orbitals Bi 6p, Pt 6s and Pt 5d. Results are shown in Fig.1 in main text.
	
	To evaluate the robustness of the Weyl nodes, we performed calculations for the trigonal crystal structure reported in Ref.~\citenum{Kaiser2014}. These confirm the presence of Weyl nodes albeit, in this case, they are found at higher energy (96 meV). Noteworthy, between these two crystal structures, the values of $a$ and $c$ differ by less than 0.1\% while differences in the Bi coordinates lead to a van der Waals gap (zvdW) 3\% smaller in our refinement. A third calculation based on an artificially enlarged zvdW in our structural model yields the Weyl nodes at 79~meV indicating that, in fact, zvdW controls to a large extent the Weyl node energy.
	
\begin{figure}[h!]
		\centering
		\includegraphics[width=0.5\textwidth]{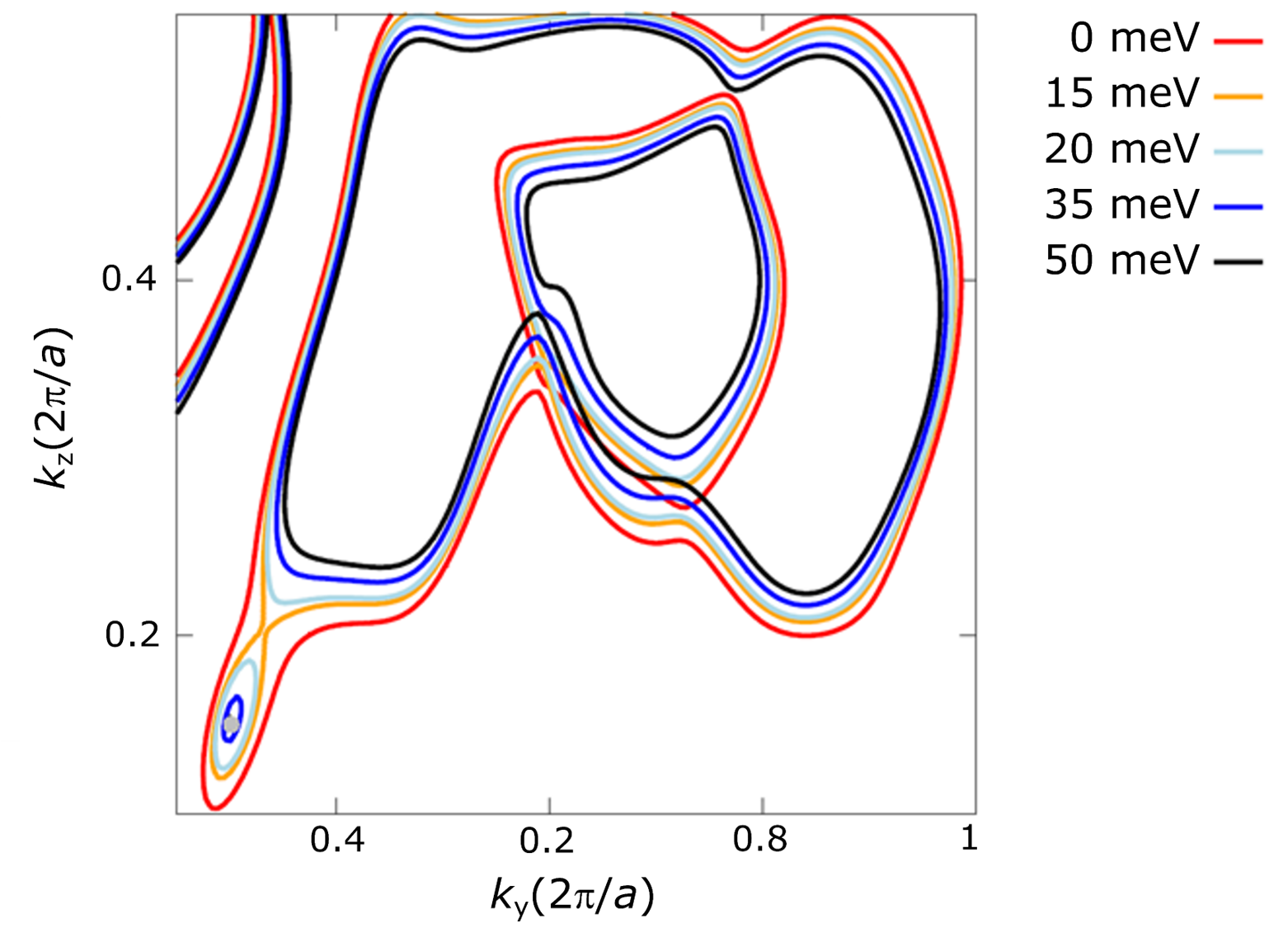}
		\caption{Isoenergetic contours of the band associated with the Weyl node. The red line corresponds to the Fermi energy.}
		\label{FigS1-2}
	\end{figure}

	Fig.~\ref{FigS1-2} shows the isoenergetic contour on a larger momentum scale than Fig.1 in the main text. Here the energy is measured with respect to the chemical potential obtained by DFT, 0~meV corresponding to the Fermi energy. The band associated with the Weyl node has a larger pocket and the two of them merge at a certain doping level.
	
\begin{figure}[h!]
		\centering
		\includegraphics[width=0.5\textwidth]{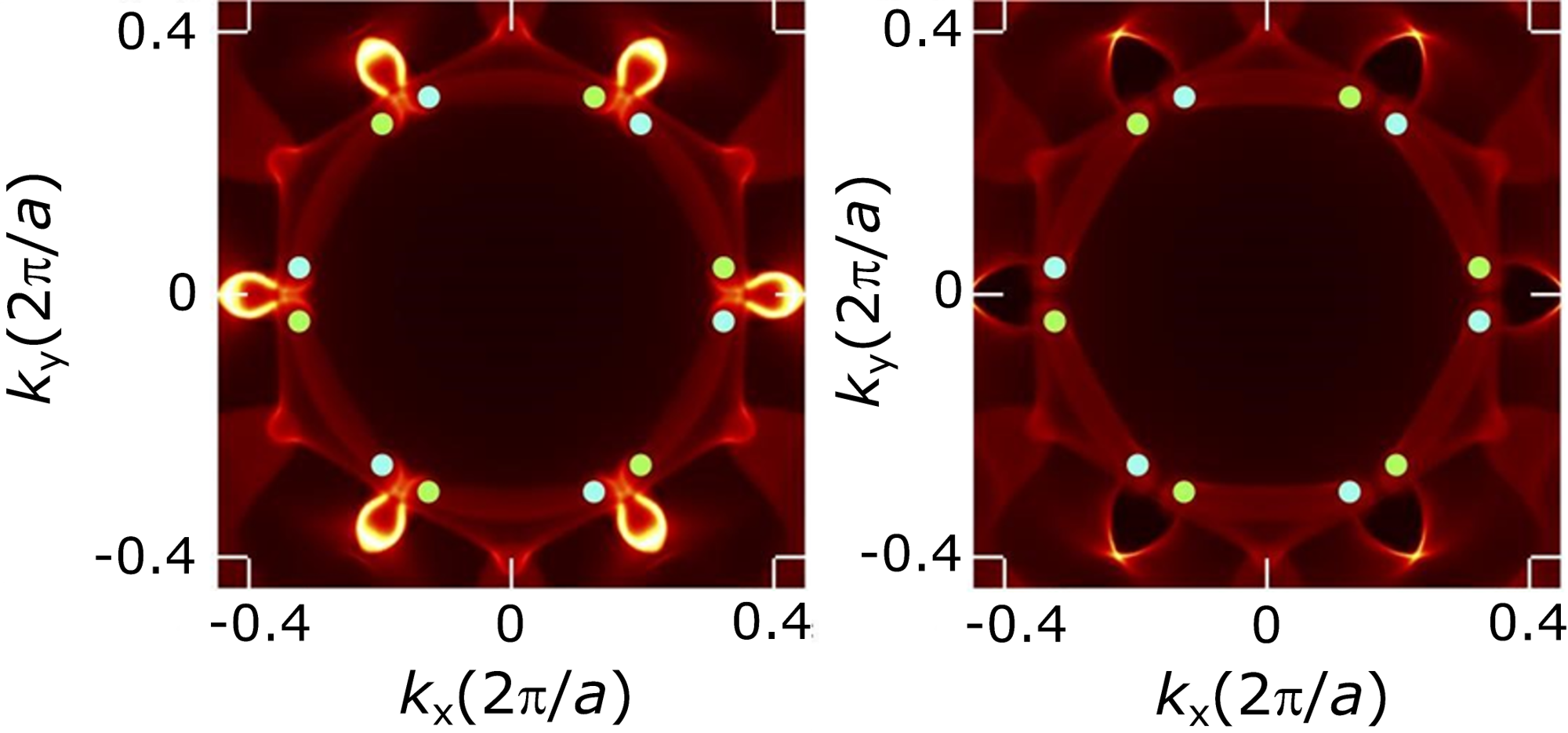}
		\caption{Surface Fermi surface corresponding to the charge neutrality point. Green (light blue) points correspond to the projection of Weyl nodes of positive (negative) chirality. Left: [001] Bi$_2$-terminated surface. Right: [001] Bi$_4$-terminated surface.}
		\label{FigS1-3}
	\end{figure}

	\textbf{Fermi arcs.} A calculation based on a semi-infinite slab along the 001 direction shows that the surface Fermi energy contours in \ch{PtBi2} present a strong sensitivity to the surface termination. In particular, clear spectral weight connecting Weyl nodes of opposite chirality can be observed, which is more intense for a \ch{Bi2}-terminated surface. Fig.~\ref{FigS1-3} evidences the presence of Fermi arcs for both Bi$_2$-terminated surfaces as well as for Bi$_4$-terminated surfaces.
	
\begin{figure}[h!]
		\centering
		\includegraphics[width=1\textwidth]{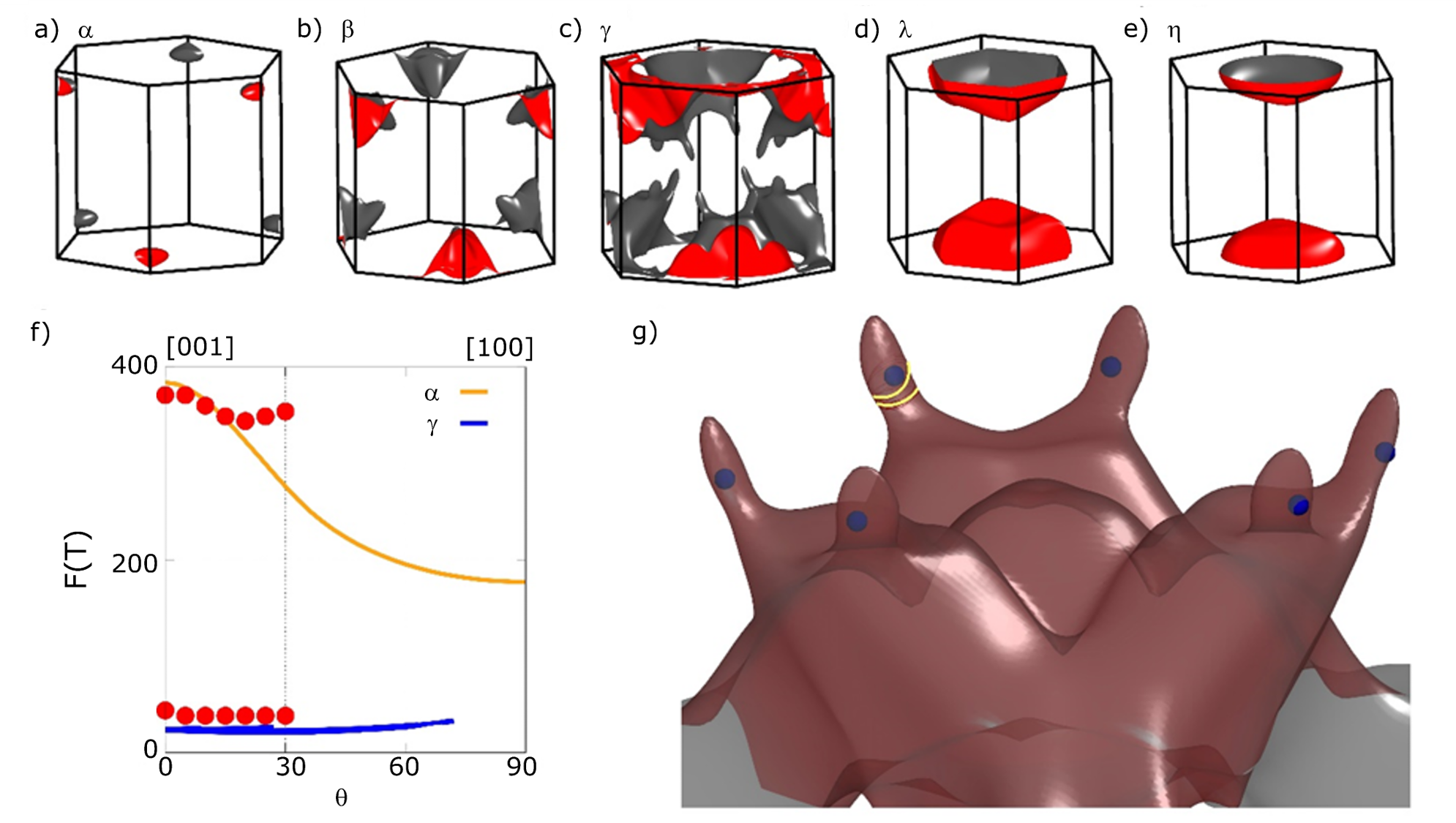}
		\caption{(a-e) Fermi surface as obtained from density-functional calculations. The outer (inner) face is colored gray (red). (f) Computed quantum-oscillations spectrum at low frequencies. Red points correspond to our experimental results. (f) Extended view of the Fermi surface $\gamma$. Weyl nodes are shown as blue points and characteristic extremal orbits of the lowest frequency branch as yellow curves.}
		\label{FigS1-4}
	\end{figure}

	\textbf{Quantum oscillations.} We computed the spectrum of quantum oscillations of \ch{PtBi2} (SG 157) starting from our density-functional calculations.  We used the de Haas van Alphen (dHvA) module of the FPLO code based on the Wannier tight-binding Hamiltonian. The starting point of this method is the calculation of Fermi surface, which is initially computed based on a mesh of the Brillouin zone having $18 \times 18 \times 18$ subdivisions. Then, the Fermi surface is refined based on an adaptive mesh. The number of refinement rounds performed in our calculations was five.  The Fermi surface obtained is shown in Fig.\ref{FigS1-4} (a-e). The main characteristics are in good agreement with Ref.~\citenum{Gao2018} and Ref.~\citenum{Shipunov2020}.
	
	As obtained in Ref.~\citenum{Gao2018}, the quantum-oscillations spectrum extends over a very large range of frequencies. Here we present results in a relatively low-frequency window, shown in Fig.\ref{FigS1-4} (f) together with the experimental data. Observed frequencies agree very well with our band structure calculations. It is at the present unclear why in our experiments the oscillations are visible only in a narrow angular range. For the field along [001], the frequency $\sim 350$~T stems from the trivial pocket named $\alpha$ in Ref.~\citenum{Gao2018}. On the other hand, the smallest frequencies originate in extremal orbits located at the finger-like parts of the pocket $\gamma$. This pocket is the one that encloses the Weyl nodes which are in fact located extremely close to such extremal orbits, as shown in Fig.\ref{FigS1-4} (g).
	
	\section{\label{TempDependence} Temperature dependence and magnetoresistance}
	
\begin{figure}[h!]
		\centering
		\includegraphics[width=0.6\textwidth]{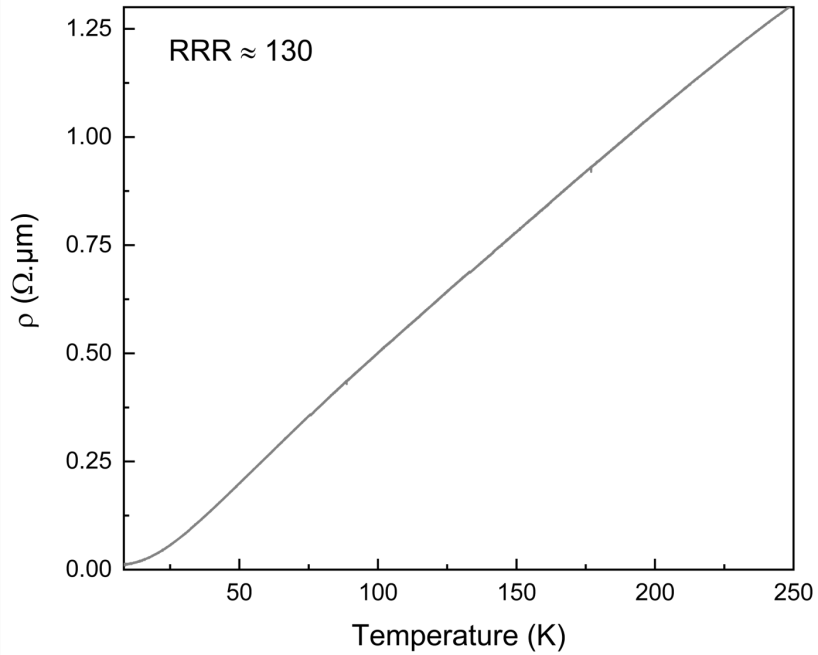}
		\caption{Temperature dependence of the resistivity of a macroscopic crystal between $T=300$~K and $T=4.2$~K.}
		\label{FigS2-1}
	\end{figure}

	\textbf{Temperature dependence.} We present in Fig.\ref{FigS2-1} below the temperature dependence of the resistivity of a single macroscopic crystal exhibiting a residual resistance ratio of about 130.
	
\begin{figure}[h!]
		\centering
		\includegraphics[width=0.6\textwidth]{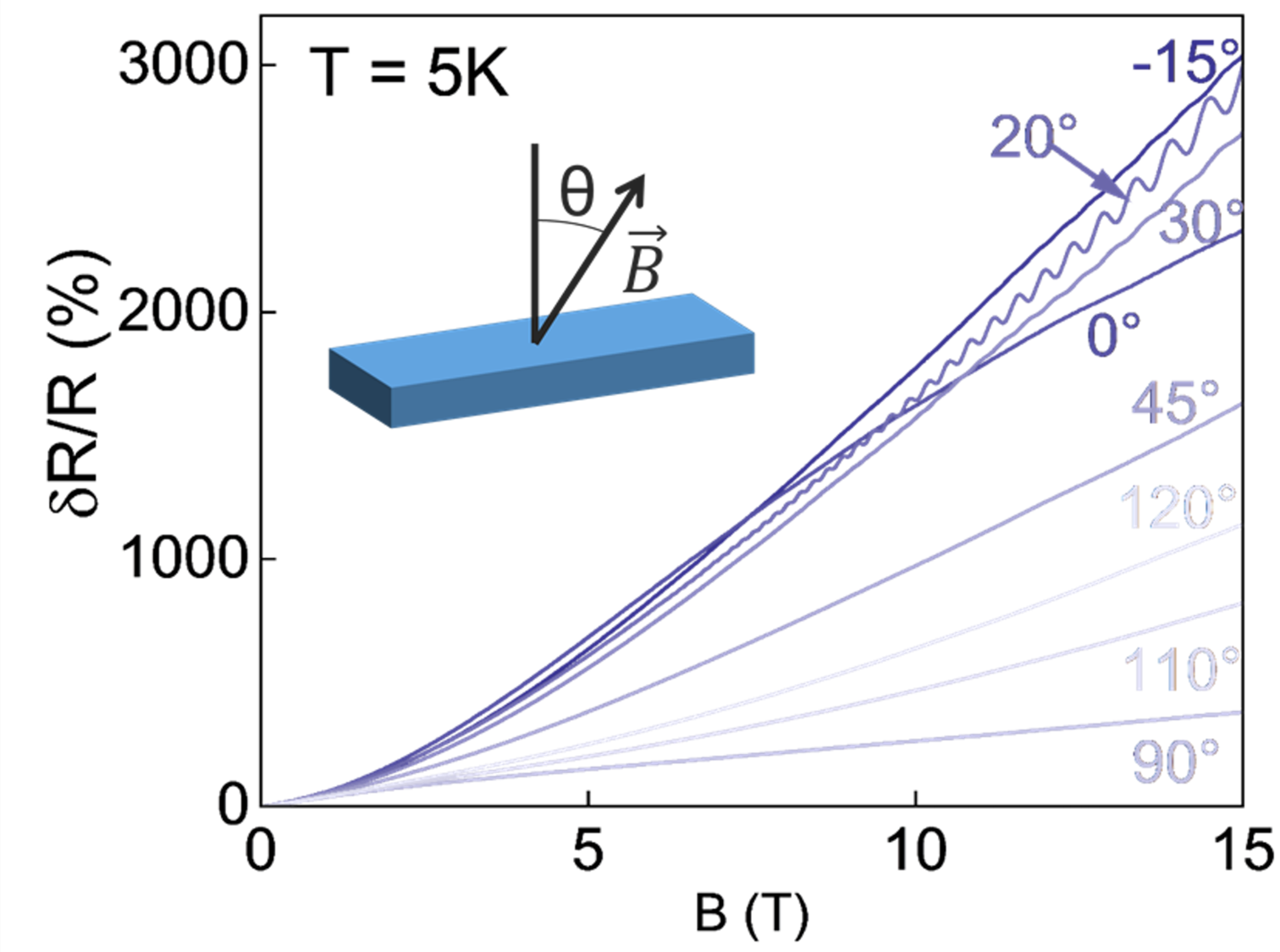}
		\caption{Magnetoresistance of a single crystal for different tilted angles $\theta$ of the field with respect to the out-of-plane direction, at 5K.}
		\label{FigS2-2}
	\end{figure}

	\textbf{Magnetoresistance at $\theta=20^\circ$.} A single macroscopic crystal was mounted on a piezo driven rotator and measured under magnetic field down to 5K. At low temperature, we measured the magnetoresistance in fields tilted between -30$^\circ$ and 120$^\circ$ with respect to the out-of-plane direction. A strong anisotropy is measured (Fig.\ref{FigS2-2}), in good agreement with previously reported results \cite{Gao2018,Xing2020}.
	
\begin{figure}[h!]
		\centering
		\includegraphics[width=0.6\textwidth]{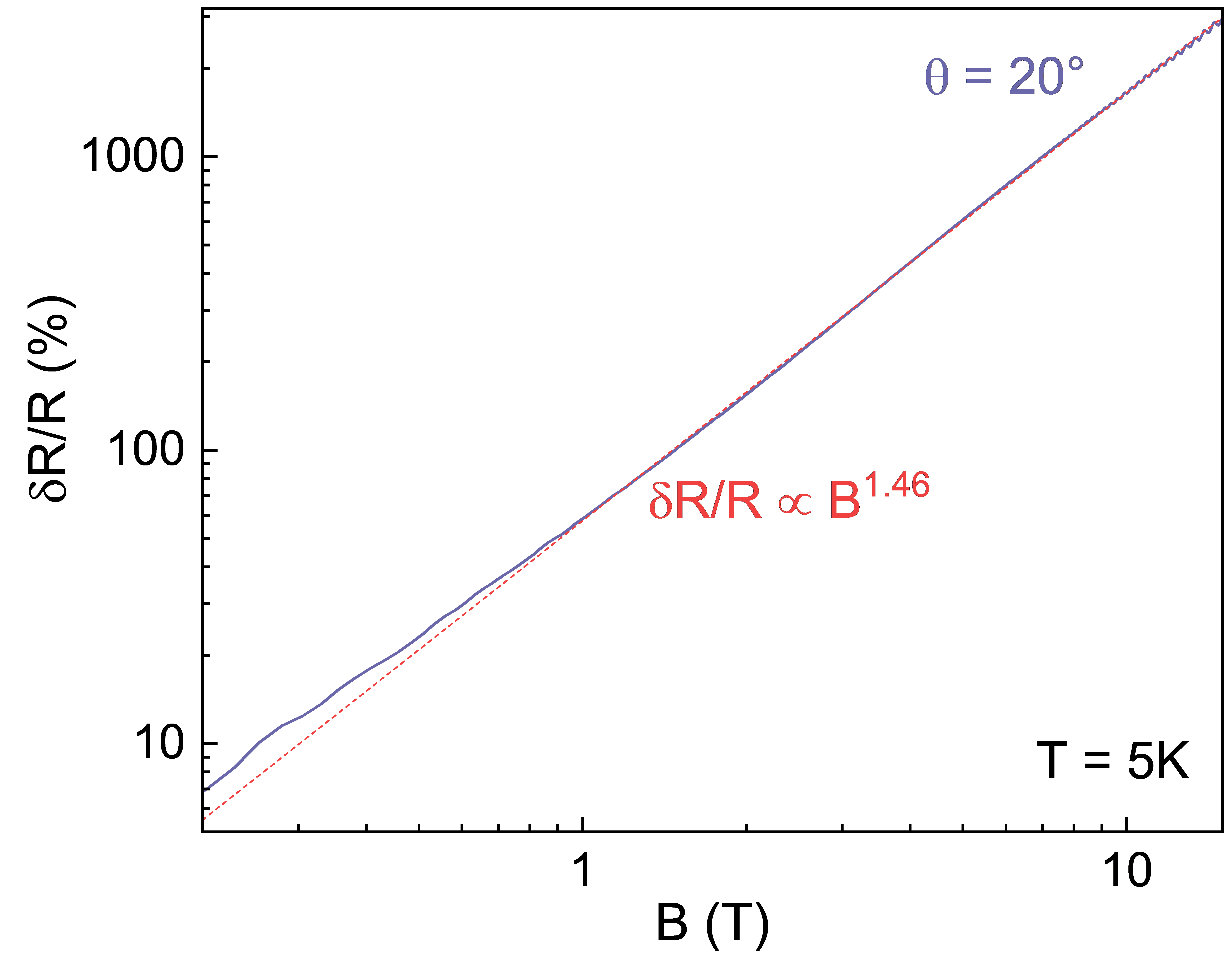}
		\caption{The magnetoresistance of the macroscopic crystal measured at $\theta=20^\circ$, $T = 5$~K and up to $B = 12$~T in a logarithm-logarithm scale. A power law behavior is evidenced between $1$~T and $12$~T with an exponent of $1.46$.}
		\label{FigS2-2a}
	\end{figure}

	The magnetoresistance at $\theta=20^\circ$ is plotted in Fig.\ref{FigS2-2a} in a logarithm scale, showing a power law dependence of $\delta R/R$ with $\delta R/R \propto B^{1.46}$ as reported in Refs.~\citenum{Gao2018} and ~\citenum{Xing2020}.
	
\begin{figure}[h!]
		\centering
		\includegraphics[width=1\textwidth]{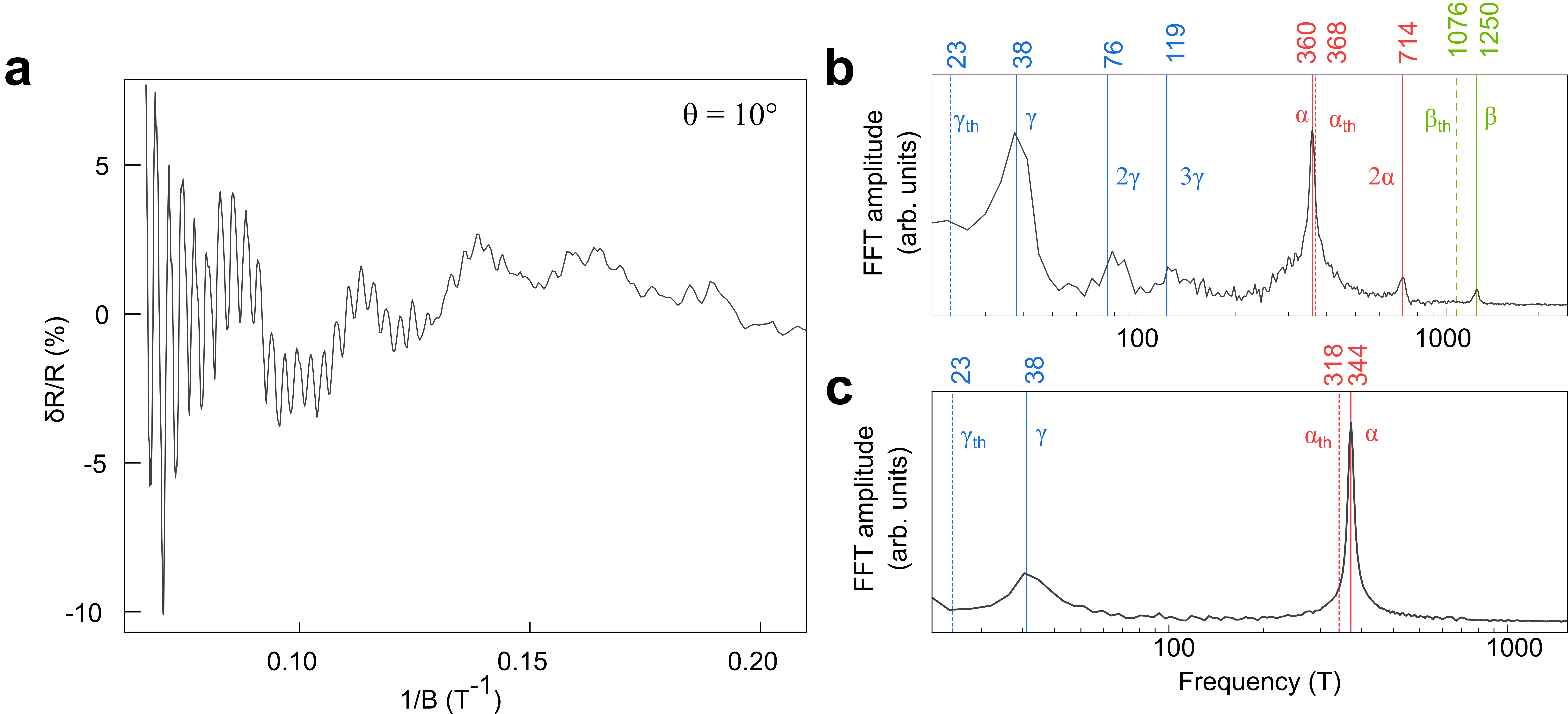}
		\caption{(a) Shubnikov-de Haas oscillations measured at $T=5$~K and $\theta=10^\circ$, obtained by removing a smooth background to the magnetoresistance. (b) Fast-Fourier Transform of the oscillations for the same data. Three peaks related to the $\alpha$, $\beta$ and $\gamma$ pockets and some of their harmonics ($2\alpha$, $2\gamma$ and $3\gamma$) could be identified. The expected frequencies from DFT calculations  $\alpha_\text{th}$, $\beta_\text{th}$ and $\gamma_\text{th}$ are indicated by the dashed lines. (c) FFT of the SdHo for $\theta = 20^\circ$ and at 5K. Vertical lines indicate the positions of the experimental (plain) and theoretically predicted (dashed) peaks.}
		\label{FigS2-3}
	\end{figure}

	\textbf{Shubnikov-de Haas oscillations.} For angles around $\theta = 20^\circ$, large Shubnikov-de Haas oscillations (SdHo) appear at low temperature, but rapidly disappear for farther angles \cite{Gao2018}. At $\theta = 20^\circ$, two peaks emerge in the fast Fourier transform of the SdHo, at 38~T and 345~T (Figure \ref{FigS2-3}.c), in very good agreement with our DFT calculation, which shows they originate from the pockets named $\gamma$ and $\alpha$ in Ref. \citenum{Gao2018}. Our calculations show that the $\gamma$ pocket is a hole pocket that encloses the Weyl nodes. It contains extremal orbits very near the Weyl nodes which yield the low-frequency quantum oscillations measured. This pocket also contains additional extremal orbits, although they are expected to contribute to frequencies much higher than those observed (see \textcolor{red}{S1}). The $\alpha$ pocket is a topologically trivial hole pocket. For $\theta = 0^\circ - 10^\circ$, additional oscillations are measured at 1250T in the Fourier transform spectrum (almost angle-independent) in agreement with some other work \cite{Gao2018} and with the $\beta$ pocket identified in our calculations (see Fig.\ref{FigS1-4}), which is a trivial pocket.
	
	The amplitude of the Shubnikov-de Haas oscillations shows some maximum at $\theta=25^\circ$ and a local maximum at $\theta=-15^\circ$ and vanishes when $\theta$ deviates from these values. The Fourier transforms (FFT) of the oscillations indicate that the presence of peaks corresponding to the $\alpha$ and the $\gamma$ bands is robust for any angle showing Shubnikov-de Haas oscillations.
	
	\section{\label{Fabrication} Sample fabrication and measurement Set-up }
	
\begin{figure}[h!]
		\centering
		\includegraphics[width=0.8\textwidth]{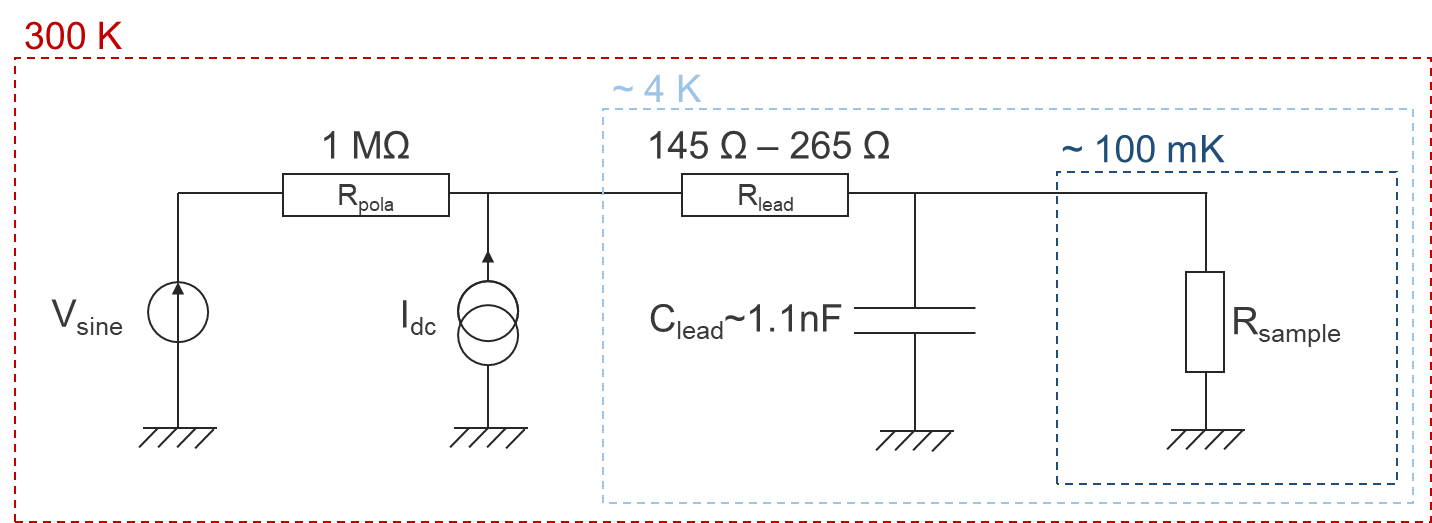}
		\caption{schematic diagram of the different electronic elements we used to measure in our dilution fridge, together with the different temperature stages.}
		\label{FigS16-1}
	\end{figure}

	\textbf{Measurement techniques.} Superconductivity in macroscopic crystals was measured using dc sources operated in a delta mode, a method that is particularly adapted to the measurement of low resistances. The measurements of nanostructures were done using standard lock-in amplifier techniques at low frequency ($f<200$~Hz). The measurement setup indicating the different filters and the connection scheme is indicated in Fig.\ref{FigS16-1}.
	
	Voltages measured at 100~mK on different sets of contact saturates at a value that is very close to but not exactly zero. This value lies typically between $+10$~nV and $-10$~nV. These values being positive as much as negative, they cannot be attributed to a residual resistance of the film. The non-vanishing voltage is the result of a measurement artefact related to common mode rejection ratio (CMRR) issues, a limitation of the measurement of low impedance samples with lock-in techniques. Indeed, the CMRR of the Lock-in amplifier we are using (Stanford Research 830) is expected to be about 100~dB but in practice, this is found to be around $85 - 90$~dB (a factor of $2.10^4-10^5$ rejection). Therefore, the value of the voltage measured in the A-B mode can fluctuate by a value given by $\pm I_\text{pola}R_\text{line}/CMRR={10}^{-6}\times250/2.{10}^4=\pm12.5$~nV, for $I_\text{pola}=1$~$\mu$A, $R_\text{line}=250\Omega$ and $CMRR=2.{10}^4$, a value that corresponds very well with our measurements on different sets of contact.
	
	\textbf{Nanostructure fabrication.} The exfoliation was made on Si/SiO$_2$ substrates, with a 290~nm oxide thickness, so as to enhance optical contrast. We did not use any glove box for the fabrication of the devices and the structures were fabricated in air. Exfoliated structures were then contacted by standard electron-beam lithography and metal lift-off to obtain ohmic CrAu contacts. The main structure studied in this work has a thickness of 60~nm and a lateral size of about 10~$\mu$m. Three other thin flakes were prepared and gave similar results (see section \ref{OtherSamples}).
	
	\section{\label{DefinitionRN} Definition and measurement of $R_\text{N}$}
	
	\textbf{Determination of the normal resistance $R_\text{N}$.} There is three different ways to extract the value of the normal resistance. All the ways are found to be equivalent in our work. $R_\text{N}$ can be defined as: (i) the resistance measured at a temperature just before the superconducting resistance drop down, when the resistance is almost temperature independent, (ii) the resistance measured when a large enough magnetic field is applied to close the gap completely or (iii) the differential resistance measured when a DC current is applied, the value of the DC current being significantly larger than the critical current and the differential resistance being almost dc-current independent.
	
	We note that it was not possible to use the method (iii) for macroscopic because for which the critical current $I_\text{c}$ was too large to measure the differential resistance at $I > I_\text{c}$ without inducing significant heating in the measurement lines.
	
	All the definitions of $R_\text{N}$ for a given set of contact are found to be perfectly equivalent in our work.
	
	\section{\label{OtherSamples} 2D superconductivity in D2, D3 and D4}
	
\begin{figure}[h!]
		\centering
		\includegraphics[width=0.8\textwidth]{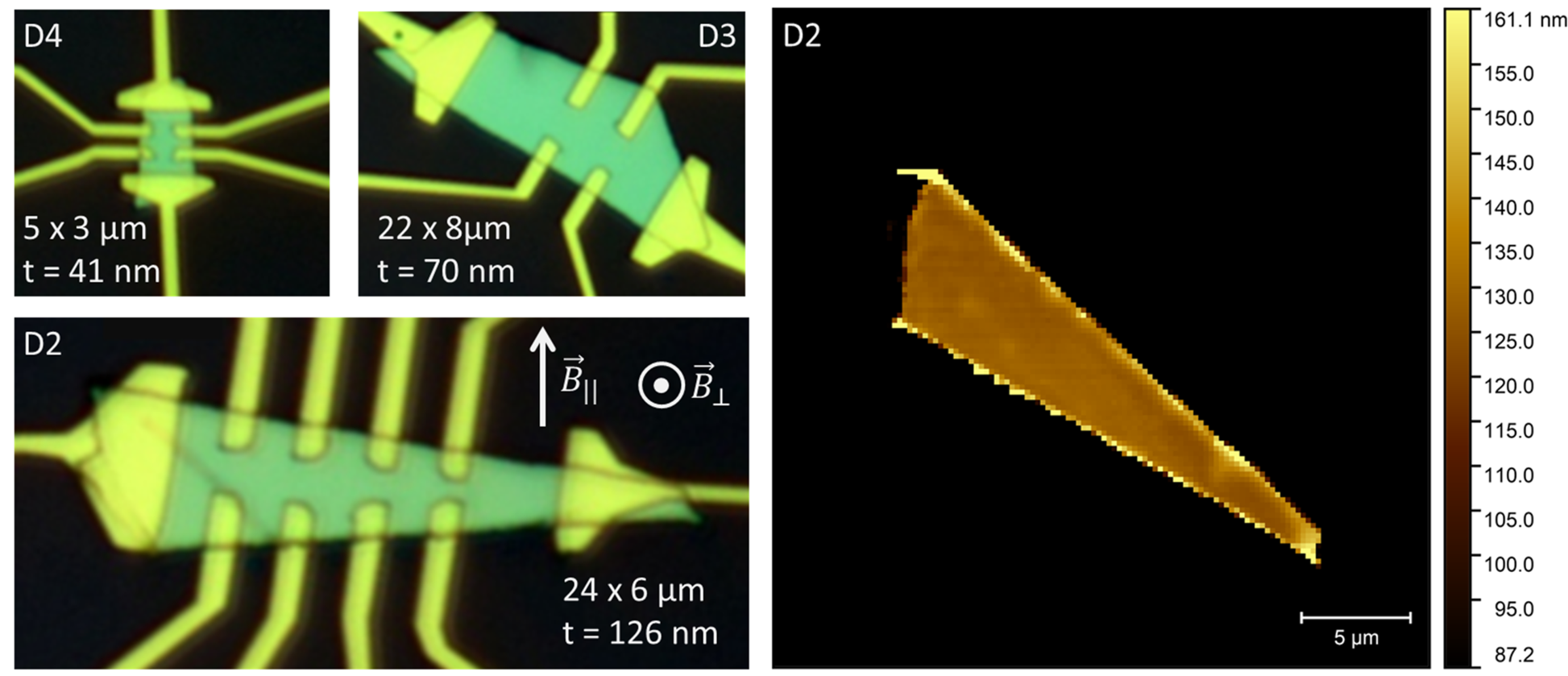}
		\caption{Left: Optical picture of the three additional nanostructures of \ch{PtBi2} with their dimension indicated in each picture and the orientation of the magnetic field we refer to in the text. All structure shows a superconducting transition. Right: Atomic force microscopy (AFM) image of D2 after exfoliation (before contacting), showing a flat surface with a similar roughness as the SiO2 substrate ($\sim1$nm). The height increase at the edges corresponds to residues of glue from the exfoliation.}
		\label{FigS7-1}
	\end{figure}

	\textbf{Devices.} Measurements on three other nanostructures (D2, D3 and D4 in Fig.\ref{FigS7-1}) are presented below with the characterization of the 2D superconducting state for different thicknesses than D1 in the main text ($t=60$~nm): one flake is significantly thicker (D2, $t=126$~nm), one flake is of similar thickness (D3, $t=70$~nm) and the last flake is thinner (D4, $t=40$~nm).
	The critical temperature $T_c$ is found to increase with the RRR as indicated in the table~\ref{Table1} with $T_c$ defined by $R(T_c)=R_0/2$ where $R_0$ is the height of the resistance step reaching the zero resistance step (he abrupt transition). This definition is generally different as the one used in the main text where $R_0$ is replaced by $R_N$ but is equivalent for very thin samples showing a single superconducting transition (D1 and D4). we also adapt below our definition of the critical magnetic field accordingly ($R(B_c) = R_0/2$). 
		
	\textbf{In-plane vs. out-of-plane critical fields.} The very low temperature ($T=100$~mK) magnetoresistances for an out-of-plane magnetic field are shown in Fig.~2 in the main text for the four nanostructures and for the macrostructures. The different out-of-plane and in-plane critical fields are summarized in the table~\ref{Table1} along with the thicknesses of the different structures, the RRR, the $T_c$ and the ratio $B_{c,\parallel}/B_{c,\perp}$.
	
	We note that $B_{c,\perp}$ and the ratio $B_{c,\parallel}/B_{c,\perp}$ depend significantly on the set of contact considered for D1. This might be due to structural inhomogeneities in the sample or to the geometry of the sample, which differs from the Hall bar type of geometry of samples D2 to D4, involving possible issues related to inhomogeneous current density in the sample. This could also explain the large value of the ratio found with respect to the other samples.
	
	\begin{table}[h!]
		\caption{In order to compare the data collected from the different devices, we report in this table the out-of-plane and in-plane critical fields, the critical temperatures and the RRR for the different structures measured for this work (D1 to D4), including the data of the structure shown in the main text (D1). The in-plane critical field considers the misalignment of the sample. The different fields indicated for D1 are related to different set of contacts.}
			\begin{tabular}{lcccccc}
				Samples & Thickness (nm) & RRR & $T_c$~(mK) & $B_{c,\perp}$~(mT) & $B_{c,\parallel}$~(mT) & $B_{c,\parallel}/B_{c,\perp}$ \\
				\hline
				D2 & 126  & 13.6 & 275 & 1.5 & 25 & 17 \\
				D3 & 70 & 8.7 & 325 & 4 & 95 & 24 \\
				D1 (main text) & 60 & 8.9 & 360 & 4-8 & 230-250 & 31-59 \\
				D4 & 41 & 4.7 & 400 & 12 & 315 & 26 \\
			\end{tabular}
		\label{Table1}
	\end{table}
	
\begin{figure}[h!]
		\centering
		\includegraphics[width=0.7\textwidth]{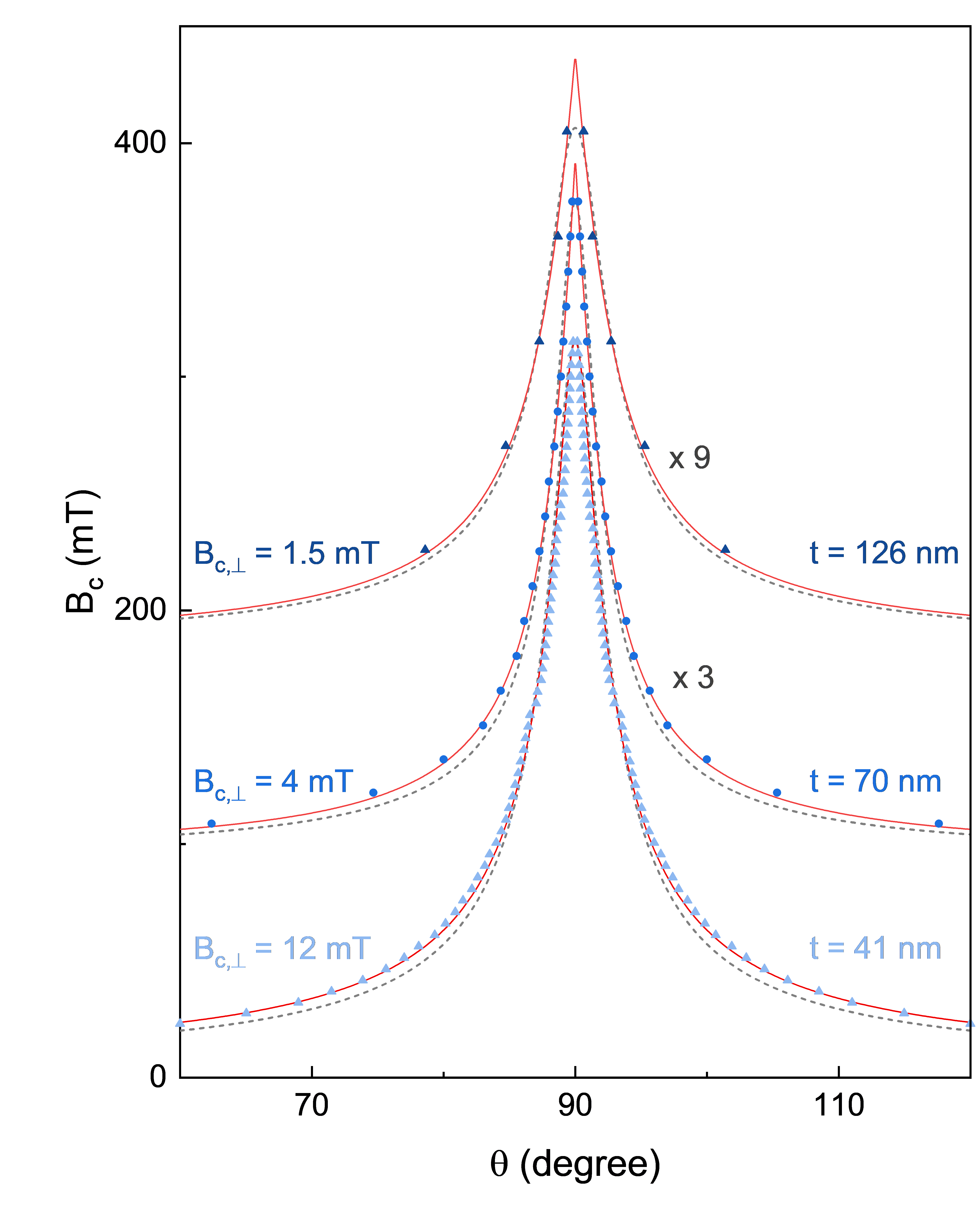}
		\caption{Angular dependence of the critical field at T=100 mK for the three samples D2, D3 and D4. It shows a sharp peak with a maximum corresponding to $\theta \sim 90^\circ$ and a misalignment corresponding to less than a few degrees. The experimental data, symmetrized along $90^\circ$ in order to get rid of any due to the remanant field of the coil, are indicated in dark blue, blue and light blue for D2, D3 and D4 respectively. For the sake of clarity, the data are shifted along the y axis (the in-plane and out-of-plane critical field are indicated in table~\ref{Table1}). The fit to the 2D Tinkham model is shown by the red line and the fit to the 3D Ginzburg-Landau model is indicated by the grey dashed line. A better agreement is found for the 2D model, even for the thicker sample D4 ($t=126$~nm).}
		\label{FigS7-3}
	\end{figure}

	\textbf{Angular dependence of $B_{c}$.} The angular dependence of the critical field for D2, D3 and D4 is shown in Fig.\ref{FigS7-3} with the fit to the 3D Ginzburg-Laudau model and to the 2D Tinkham model. Better agreement could be found with the Tinkham model, indicating the 2D nature of the superconducting state even for the thicker sample (D2, $t= 126$~nm).
	
	\section{\label{RLowT} Superconducting transition and isotropy of $B_\text{c}$ in a macroscopic crystal}
	
\begin{figure}[h!]
		\centering
		\includegraphics[width=0.6\textwidth]{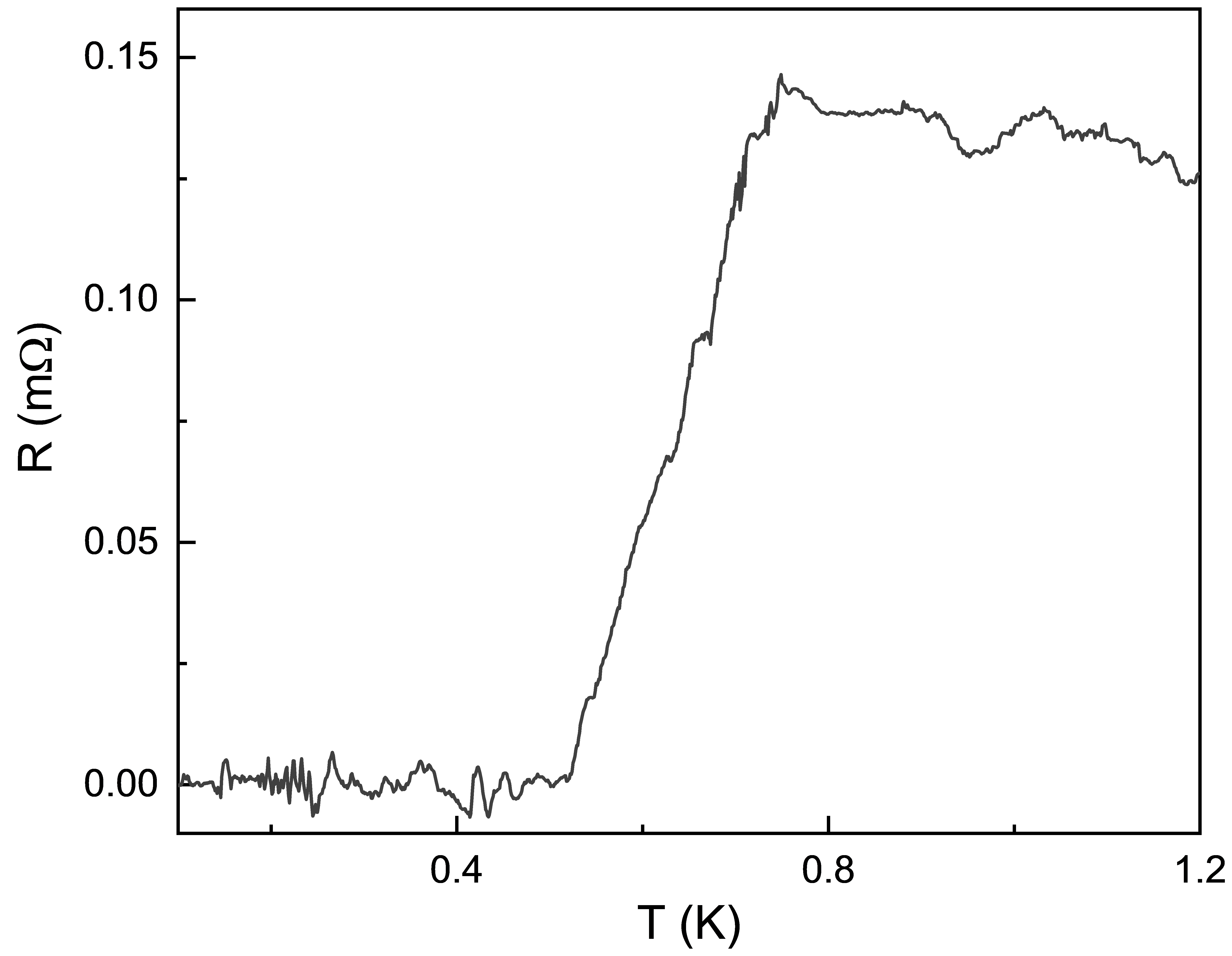}
		\caption{Low temperature dependence of the resistance of the macroscopic crystal, focusing on the temperature range close to the critical temperature. In order to increase the signal-to-noise ratio, the data are averaged by a smooth function.}
		\label{FigS4-1a}
	\end{figure}

	\textbf{Temperature dependence of the resistance.} We present in Fig.~\ref{FigS4-1a} a zoom-in of the temperature dependence of the macroscopic crystal, focusing on the superconducting transition at very low temperature.
	
\begin{figure}[h!]
		\centering
		\includegraphics[width=0.6\textwidth]{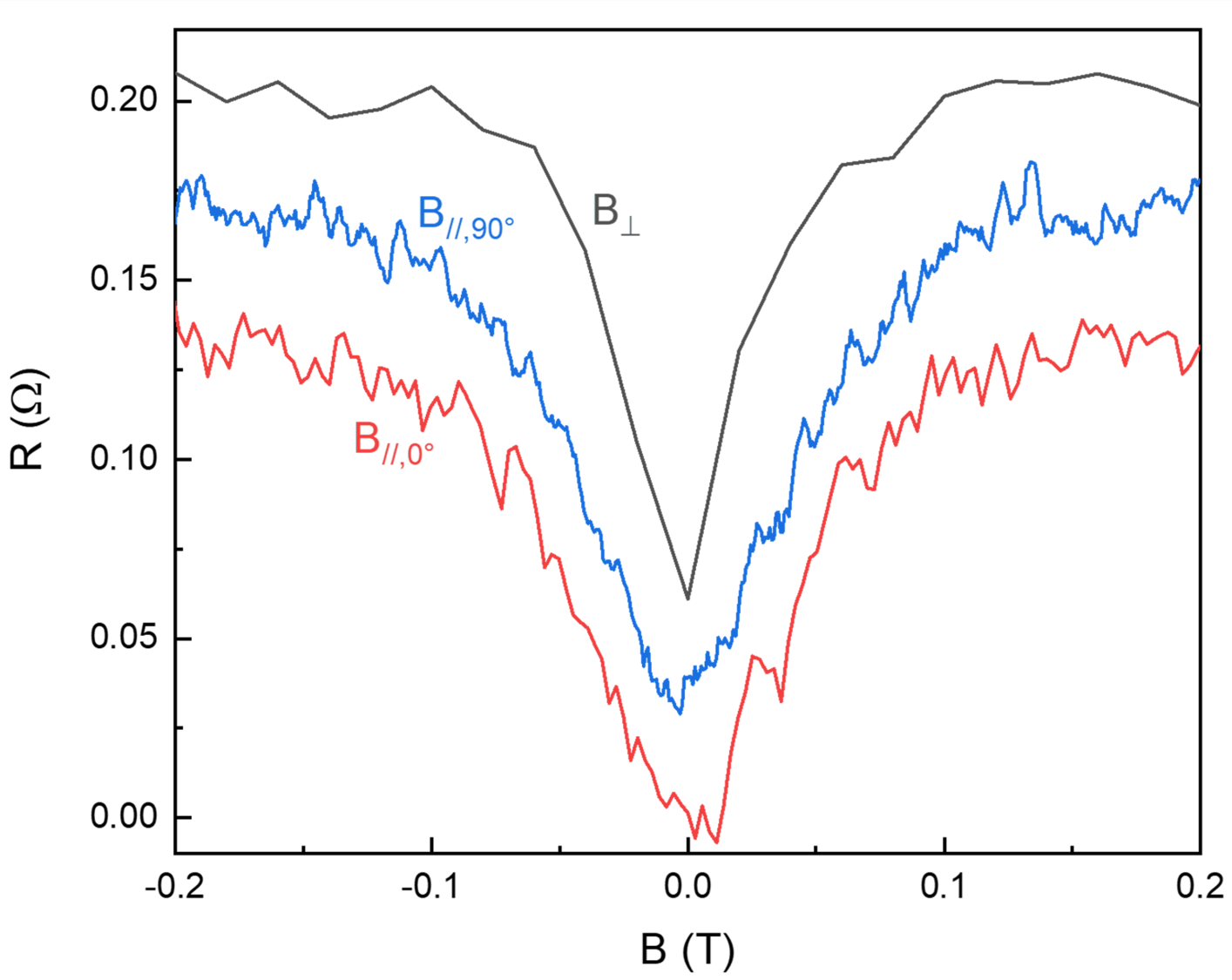}
		\caption{Magnetoresistance of a macroscopic crystal of \ch{PtBi2} with the magnetic field applied out-of-plane ($B_\perp$) or along the two perpendicular in-plane directions ($B_{\parallel,0^\circ}$ and $B_{\parallel,90^\circ}$) measured at low temperature ($T \sim 100$~mK). For the sake of clarity, the different curves are shifted in resistance.}
		\label{FigS4-1}
	\end{figure}

	\textbf{Isotropy of $B_\text{c}$.} The magnetoresistance $R(B)$ of a macroscopic single crystal of \ch{PtBi2} was measured at very low temperature along the out-of-plane direction as well as along the two perpendicular in-plane directions (Fig.~\ref{FigS4-1}). All magnetoresistances show very similar broad superconducting transition, leading to isotropic values of the critical fields $B_\text{c}$.
	
	\section{\label{MRnano} Magnetoresistances of the nanostructures}
	
	\begin{figure}[h!]
		\centering
		\includegraphics[width=0.7\textwidth]{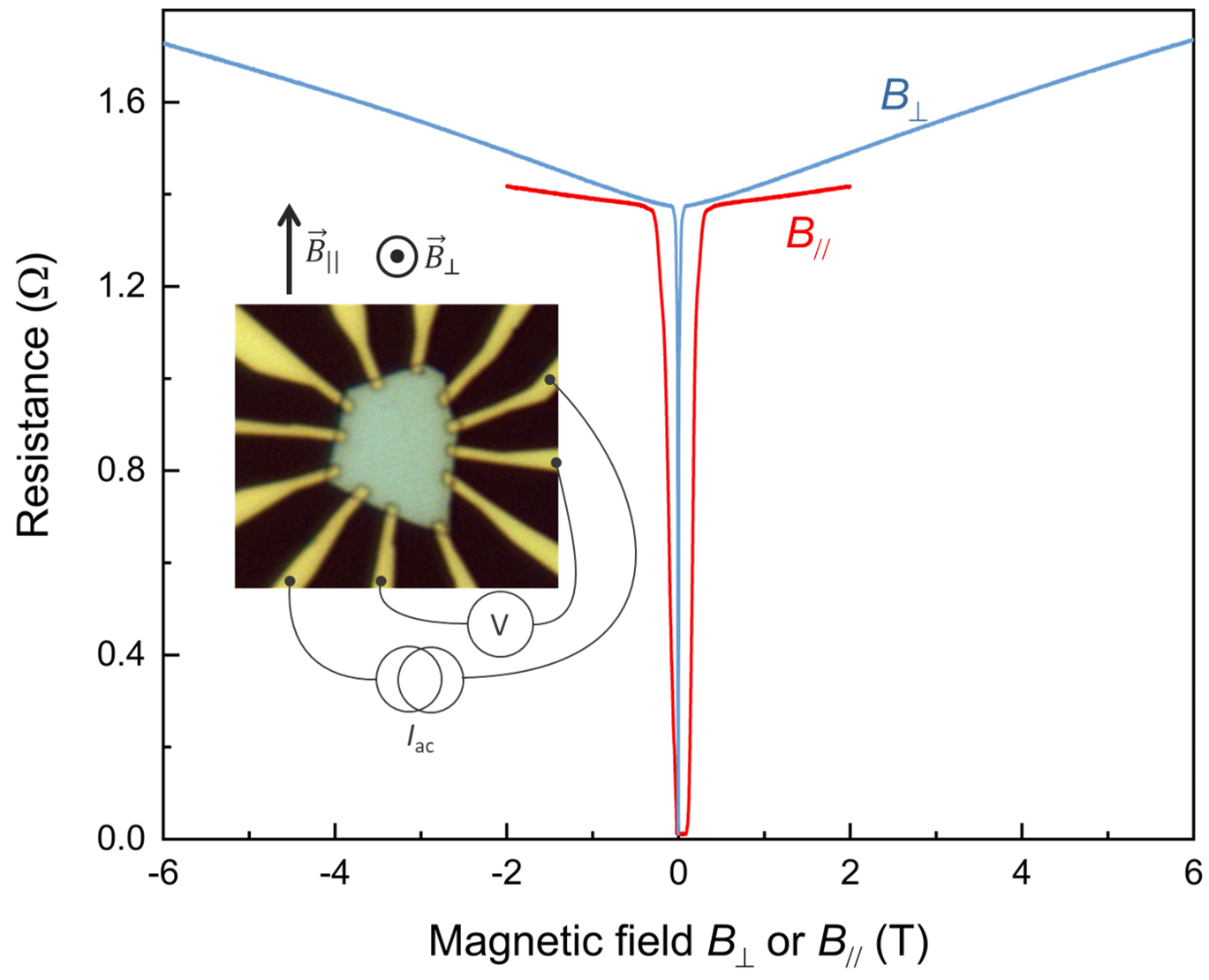}
		\caption{The magnetoresistances at $T\sim 100$~mK for an in-plane and out-of-plane magnetic fields are shown in red and blue respectively. A picture of the sample can be seen in the graph together with the connection configuration for the measurements of the magnetoresistances. The typical size of the sample is about 10~$\mu$m.}
		\label{FigS11-1}
	\end{figure}
	Fig.\ref{FigS11-1} shows the magnetoresistances along the c-axis of the exfoliated crystal and for a field aligned in the a-b plane of the nanostructure. Measurements are done at $T\sim 100$~mK and with an AC current of 5~$\mu$A. The perpendicular field can be swept up to $\pm 6$~T (the main coil of the 3D magnet system) whereas the parallel field can be swept only up to $\pm 2$~T. As for the macroscopic single crystal, the magnetoresistance is found to be almost linear and it is larger for a perpendicular field. We note that the asymmetry of the gap in the $R(B_\parallel)$ is the result of the metastable-like nature of the superconducting state, an issue that goes beyond the scope of this work.
	
	\section{\label{Bc_vs_T} Measurement of $B_{\text{c},\parallel}(T)$ and $B_{\text{c},\perp}(T)$ for a nanostructure}
	
\begin{figure}[h!]
		\centering
		\includegraphics[width=1\textwidth]{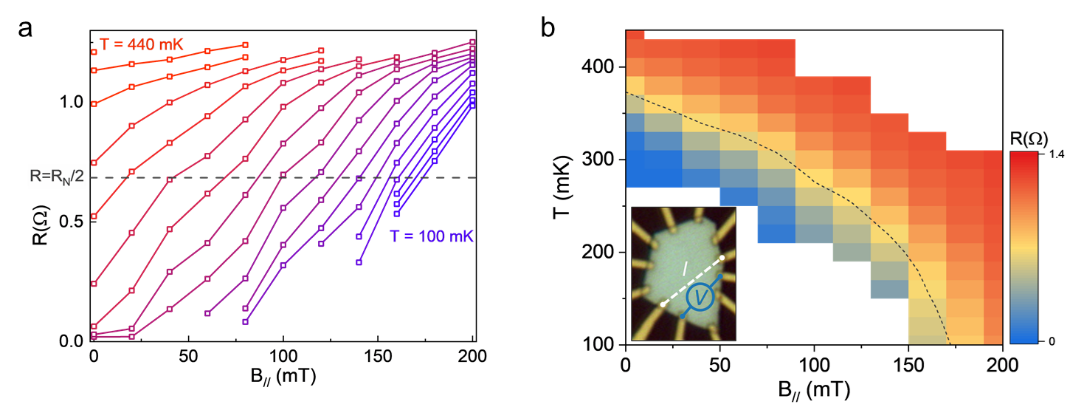}
		\caption{a) The different temperature dependence of the resistance $R(T)$ taken for different in-plane magnetic field $B_\parallel$ for a temperature ranging between 100~mK and 440~mK with steps of 20~mK. The dashed line indicates the value of half the resistance in the normal state. The dashed line crosses the different $R(T)$ at a magnetic field corresponding to the critical magnetic field $B_{\text{c},\parallel}(T)$. (b) the same data represented in a 2D color plot. The dashed line represents the temperature dependence $B_{\text{c},\parallel}(T)$. The set of contacts measured is indicated in the graph.}
		\label{FigS12-1}
	\end{figure}

	\textbf{Temperature dependence of $B_{c,\parallel}$.} In order to get the temperature dependence of $B_{\text{c},\parallel}$ and to avoid any issue due to the metastability mentioned above in section \ref{MRnano}, we swept the temperature at fixed magnetic field. Therefore, we fixed magnetic field at $0$~T first and changed the temperature by step of $20$~mK between $260$~mK and 440~mK (the range depends on the magnetic field as it can be seen in the figure below). For each step, we stabilized the temperature during more than 30~min. We finally average the value obtain over the last 50~s, when the temperature is very stable, in order to reduce the noise level.  The temperature is then increased by 20~mK to acquire the next temperature point until we reach 440~mK. $B_\parallel$ is then increased by 20~mT as we let the temperature go back to the base temperature. We repeat this previous procedure until the magnetic field reaches 200~mT (Fig.\ref{FigS12-1}).
	
\begin{figure}[h!]
		\centering
		\includegraphics[width=1\textwidth]{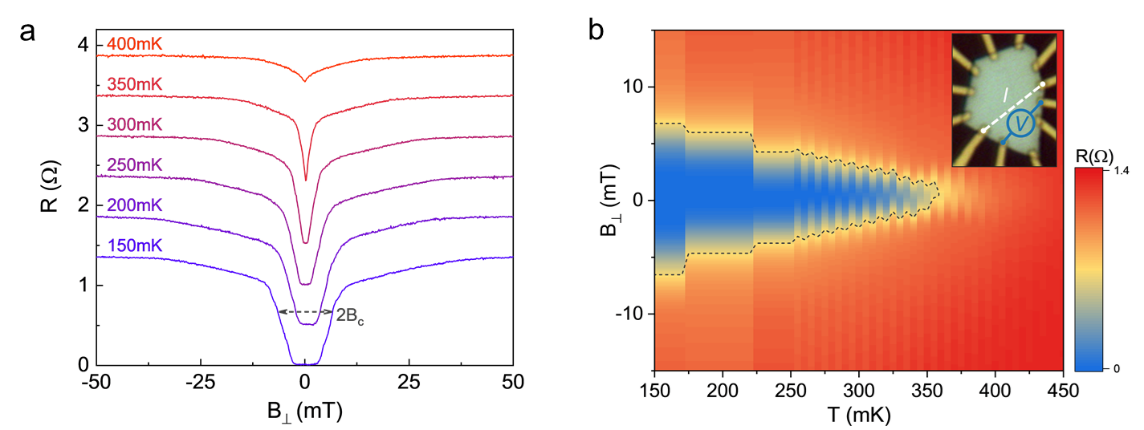}
		\caption{(a) Magnetoresistance sweeps $R(B_\perp)$ at different temperature. For the sake of clarity, the different traces are shifted and we only show in the upper graph the magnetoresistance at every 50~mK. The critical field is defined as half of the width of the gap at $R=R_\text{N}/2$ as shown by the double arrow in the $T = 150$~mK sweep. (b) the same data including the sweeps at all the temperatures and plotted in a 2D color plot. The dashed line represents the temperature dependence $B_{\text{c},\perp}(T))$. The set of contacts measured is indicated in the graph.}
		\label{FigS12-2}
	\end{figure}

	\textbf{Temperature dependence of $B_{c,\perp}$.} The procedure for the determination of $B_{\text{c},\perp}(T)$ is different due to the absence of any metastability in the out-of-plane direction. In this case, the magnetic field is swept at a fixed temperature that have been previously stabilized. The field is swept either from positive to negative field or from negative to positive fields, inducing some shift between two consecutive sweeps due to the remanant field of the coil, and the critical field is taken as half of the width of the gap (Fig.\ref{FigS12-2}).

	\section{\label{Bc_vs_ThetaAndT} Determination of $B_{\text{c}}(\theta, T)$ }
	
	\textbf{Temperature dependence of $B_{c,\parallel}$ for D1.} In order to consider any misalignment of 2D flake with respect to the different axis of the vector field and we derive the expected temperature dependence of $B_\text{c}$ for any $\theta$ angle with $\theta$ the angle of the magnetic field with the out-of-plane direction. Therefore, we replace in the solution of the equation (4) in the main text, the thermal dependence of $B_{\text{c},\perp}(T)$ and $B_{\text{c},\parallel}(T)$ given in the main text by equation (1) and (2) respectively. Such a solution is given by:
	\begin{equation}\label{Bc(theta)}
		B_\mathrm{c}\left(\theta,T\right)=\frac{1}{2}\frac{B_{\text{c},\parallel}^2}{B_{\text{c},\perp}\left(T\right)}\frac{\left|\mathrm{cos}\theta\right|}{{\mathrm{sin}}^2\theta}\left(\sqrt{1+4\frac{B_{\text{c},\perp}(T)^2}{B_{\text{c},\parallel}(T)^2}\frac{{\mathrm{sin}}^2\theta}{{\mathrm{cos}}^2\theta}}-1\right).
	\end{equation}
	Even if this solution is very close to the perfectly aligned flake for $\theta \sim 0^\circ$, large deviations can be observed close to $\theta = 90^\circ$. Consequently, we use this formula with $\theta = 91.1^\circ$ to fit the experimental data of D1 in the main text.
	
	\textbf{Temperature dependence of $B_{c,\parallel}$ for D2, D3 and D4.} The formula (\ref{Bc(theta)}) above is obtained in the framework of the Ginzburg-Landau theory and is therefore valid only for temperatures close to the superconducting transition. If a very good agreement is obtained for D1 over the full temperature range, it is not the case for D2, D3 and D4.
	
\begin{figure}[h!]
		\centering
		\includegraphics[width=0.6\textwidth]{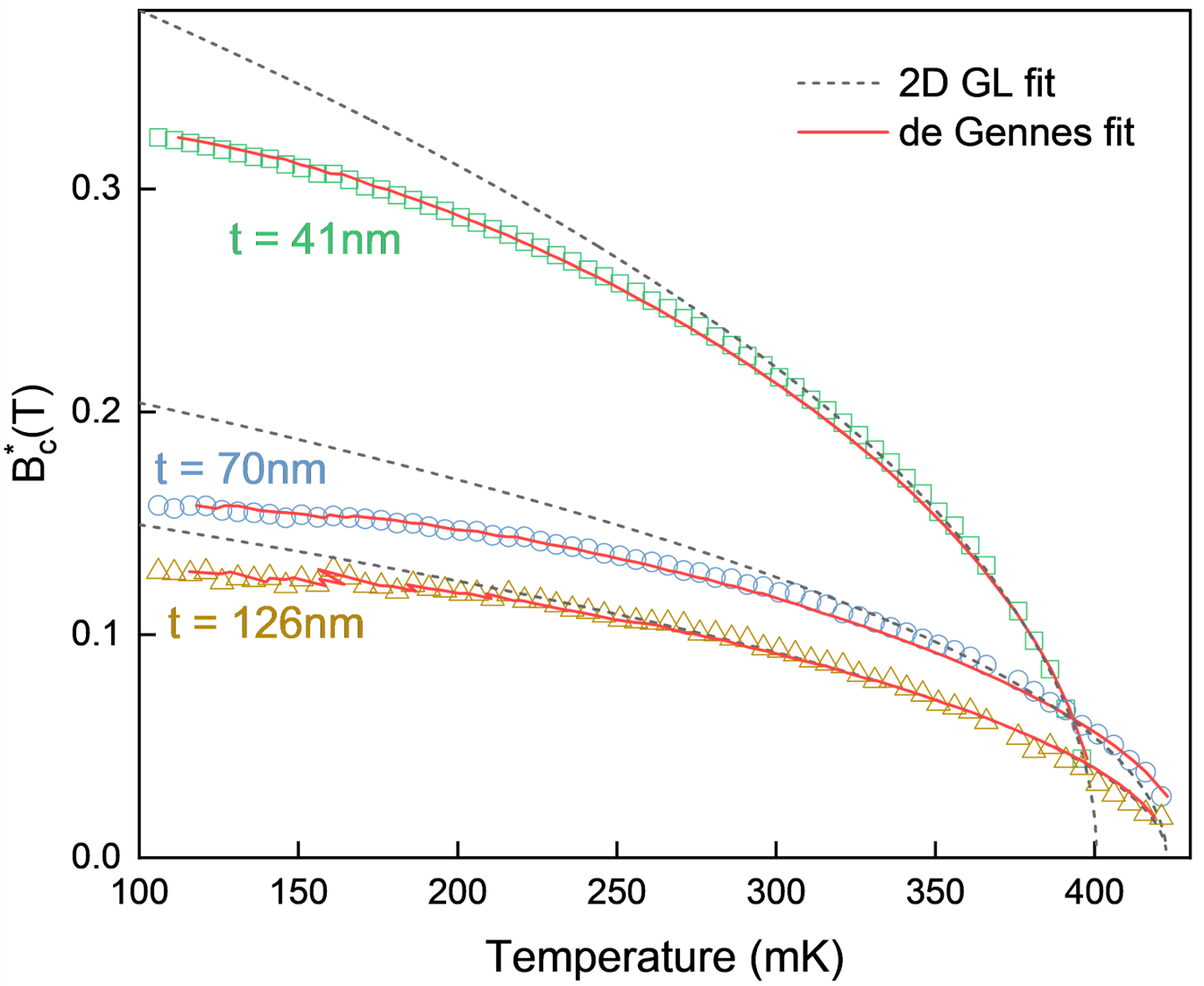}
		\caption{Temperature dependence of the in-plane critical field for the devices D2 ($t= 126$~nm in yellow), D3 ($t= 71$~nm in blue) and D4 ($t= 41$~nm in green). $B_c^*$ corresponds here to the in-plane field such that $R(B_c^*)=R_N/2$ with $R_N$ is the resistance in the normal state. The fit to the Eq.~(2) of the main text is indicated by the grey dashed lines. It is realized for temperature close to $T_\text{c}$ for reasonable agreement with the experimental data. The thickness is considered as a fixed parameters (126~nm, 70~nm and 41~nm respectively) whereas the misalignment angle is a free parameter. The results of the fits give $\xi_\parallel=53$~nm, 70~nm and 126~nm for D2, D3 and D4 respectively. The fit to the temperature dependence of $B_{c,\parallel}(T)$ given by Eq.~(\ref{PairBreaker}) above is shown in red plain lines. We found $C_1=0$~mK.T$^{-1}$ for evry fit, $C_2=$2880~mK$^2$.T$^{-1}$, 3600~mK$^2$.T$^{-1}$, 443~mK$^2$.T$^{-1}$ and $T_\text{c}=$423~mK, 319~mK and 402~mK for D2, D3 and D4 respectively.}
		\label{FigS14-1}
	\end{figure}

	For these devices, providing the assumption of an ergodic system with $\ell \ll \xi$ (which is reasonable in our case thanks to the very long superconducting coherence length $\xi$), it is possible to obtain the temperature dependence of $B_\text{c}$ over the full range of temperature. In such a regime and in the presence of a pair-breaking mechanism such as an external magnetic field, magnetic impurities or a spin-orbit coupling, the superconducting transition temperature $T$ is reduced following a universal behavior:
	\begin{equation}\label{PairBreaker}
		\ln \frac{T}{T_\text{c}} = \psi \left(\frac{1}{2}\right)-\psi \left(\frac{1}{2}+\frac{\alpha}{2\pi k_\text{B}T}\right),
	\end{equation}
	where $\psi$ is the digamma function, $k_\text{B}$ is the Boltzmann constant, $T_\text{0}$ is the superconducting transition temperature whitout any pair-breaking perturbation and $\alpha$ is the pair-breaking energy associated with the perturbation \cite{Gennes1966,Tinkham1975}. When several mechanisms are involved in breaking the superconductivity, it is possible to take all of them into account by summing their associated $\alpha$ in Eq.~(\ref{PairBreaker}). In our case, both an external magnetic field and a spin-orbit coupling can play the role of pair-breaking perturbation. According to Ref.~\citenum{Tinkham1975}, an in-plane magnetic field leads to  $\alpha_\parallel \propto B^2$ whereas an out-of-plane field leads to $\alpha_\perp \propto B$. The spin-orbit coupling also leads to $\alpha_\text{so} \propto B^2$. Therefore, we fit the temperature dependence of $B_{\text{c},\parallel}$ with the Eq.(\ref{PairBreaker}) taking $\alpha=2\pi k_\text{B}C_1 B + 2\pi k_\text{B}C_2 B^2$ with $C_1$ and $C_2$ some positive fitting parameters. We note that $C_2$ stands for both the in-plane magnetic field and the spin-orbit coupling. The results presented in Fig.~\ref{FigS14-1} are in very good agreement with the experimental data. For every sample, best fits are obtained for $C_1=0$, suggesting a weak influence of the misalignemnt in the temperature dependence of $B_{\text{c},\parallel}$ for D2, D3 and D4.

	\section{\label{R_T_OtherContact} R(T) for different contacts in D1 and BKT transition in D4}
	
\begin{figure}[h!]
		\centering
		\includegraphics[width=1\textwidth]{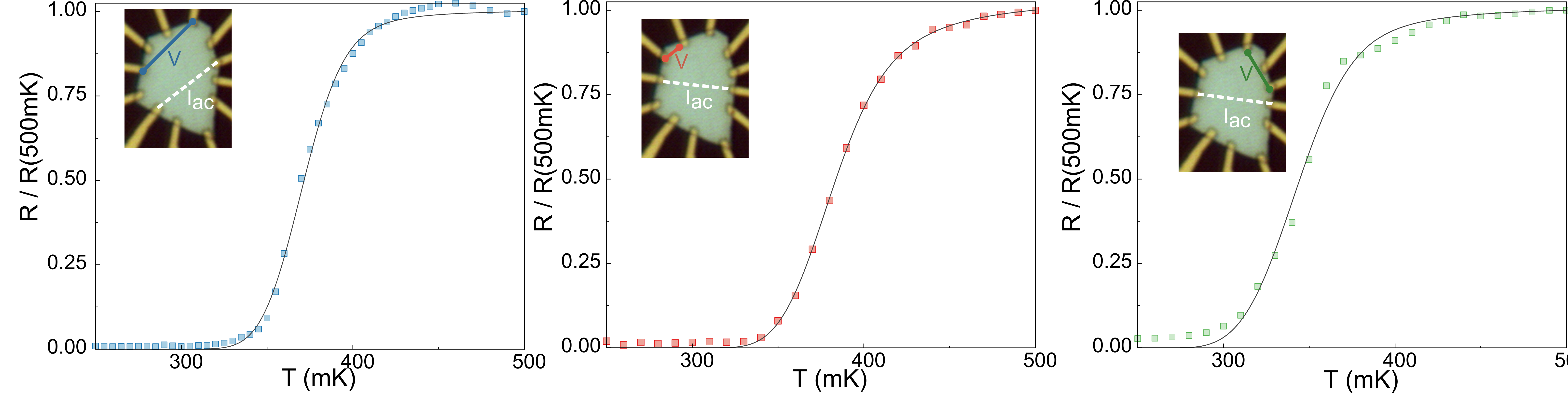}
		\caption{The $R(T)$ measured for different set of contacts indicated in the inset and fitted by equation 5 in the main text. The ac current was 1~$\mu$A and the measurements were done by increasing the temperature by steps of 10~mK with a total stabilization time longer than one hour so that the sample is very well thermalized.}
		\label{FigS15-1}
	\end{figure}

	\textbf{BKT transition for different contact in D1.} Three additional temperature dependences were measured and fitted with the BKT model for different sets of contacts for voltage probes, source and drain in D1. The results are shown in the Fig.\ref{FigS15-1}.
	
	We summarize in table~\ref{Table2} the results obtained from the fit of the different experimental data with the equation (7) in the main text. The value of the BCS critical temperature is given here by $R(T_\text{c})=R_\text{N}/2$ whereas the value of $T_\text{BKT}$ is fixed by the analysis of the $I(V)$ at different temperature as explained in the main text. The value of $\alpha$ is calculated based on the parameter given by the fit of $R(T)$.
	\begin{table}
		\caption{Results of the temperature dependence of the resistance from the normal regime down to below the BKT transition. The value of $T_{\mathrm{BKT}}$ is fixed by non-equilibrium measurements and $T_\mathrm{c}$ is given by the condition $R(T_\text{c})=R_\text{N}/2$ whereas $b$, $A$ and $\delta$ are the free parameters of the fit. $\alpha$ is calculated based on the value of $b$, $T_{\mathrm{BKT}}$ and $T_\mathrm{c}$.}
			\begin{tabular}{lcccccc}
				contact & $T_\text{BKT}$ (mK) & $T_\text{c}$ (mK) & parameter $b$ & parameter $A$ & $\alpha$ & $\delta$ (mK) \\
				\hline
				main text & 310  & 370 & 1.15 & 13.4 & 1.3 & 23.6\\
				blue & 345 & 372 & 1.09 & 57.8 & 1.9 & 15.1\\
				red (main text) & 340 & 383 & 1.14 & 25.1 & 1.6 & 16.1\\
				green & 310 & 346 & 1.10 & 30.9 & 1.6 & 19.5\\
			\end{tabular}
		\label{Table2}
	\end{table}
	Remarkably, the standard deviation of $T_\text{BKT}$ is in very good agreement with the typical value obtained for $\delta$. The small disagreement between the fit and the data in the low resistance regime is due to common mode rejection ratio as explained in the section \ref{Fabrication} above. Finally, we notice that for other sets of contact, the upturn in $R(T)$ at temperature close to $T_\text{c}$ due to inhomogeneous superconductivity prevents the fit of $R(T)$ with the BKT theory. An example of a weak upturn can be found in the blue sample in Fig.\ref{FigS15-1}.
	
	Interestingly, as shown in table~\ref{Table2}, the values of $\alpha$ given by the BKT fit of $R(T)$ are found to be systematically above the value expected for the standard XY model (up to 1.9), which predicts $\alpha<1$. According to the BCS theory, $\alpha$ is expected to be $ \sim 1/\pi$ \cite{Hu1972,Alama1999}. However, larger values of $\alpha>1$ are calculated \cite{Benfatto2007, Benfatto2008} and measured \cite{Baity2016} in high temperature superconductors like cuprates.
	
\begin{figure}[h!]
		\centering
		\includegraphics[width=1\textwidth]{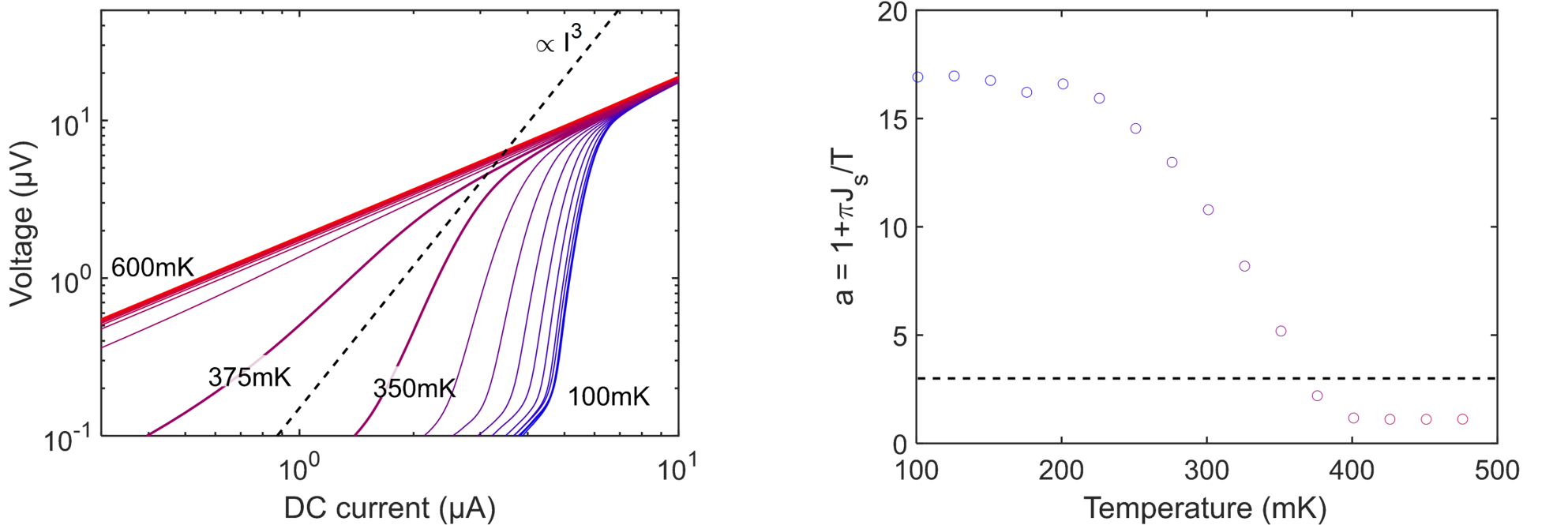}
		\caption{left: V(I) characteristics for different temperatures, in logarithmic scale. The dashed line stands for $V \propto I^3$. right: Temperature dependence of exponent a(T), with $a=3$ represented by a dashed line.}
		\label{FigS15-2}
	\end{figure}

	We finally present in Fig.\ref{FigS15-2} the $I(V)$ measurements realized on D4 ($t=41$~nm), a sample that also shows a BKT transition as indicated in the Fig.~4 of the main text.
	
	Similarly to the measurements done in D1, we evidence a cubic dependence of the $V(I)$ for $T$ between $350$~mK and $375$~mK (left in Fig.\ref{FigS15-2}). Ploting the evolution of the exponent with the temperature (right in Fig.\ref{FigS15-2}), we can extrapolate a BKT transition temperature of about $T_\text{BKT} \sim 370$~mK. As shown in the main text, the $R(T)$ can be fitted by the homogeneous Halperin-Nelson model or with the inhomogeneous model introduced by Benfatto et al. \cite{Benfatto2009} with $T_\text{BKT} \sim 370$~mK and $\delta=12$~mK $\pm7$~mK. The error bar in the determination of $\delta$ is about the same order a magnitude than $\delta$ itself which points at a very weak effect of the broadening. 
	
	Importantly, the significantly smaller value of $\delta$ found for D2 with respect to what have been measured in D1 rules out any heating by radio frequency radiation effect as a source of the broadening of the transition since both measurements were done with the same experimental setup.
	
	If the presence of inhomogeneities influences the BKT transition and more particularly the $R\left(T\right)$ dependence, it is not the only manifestation. Other features also point at inhomogeneous properties and confirm our interpretation: (i) the presence of multiple peaks in the $dV/dI$ characteristic; (ii) the upturn of $R(T)$ just above the superconducting transition ($T \gtrsim T_c$) measured on some contact pairs (see e.g. blue curve Fig.~\ref{FigS15-1}) and that is attributed to anisotropies or inhomogeneities of the superconducting transition \cite{Vaglio1993}; (iii) the variations of the different superconducting parameters with the sample and the set of contacts considered, as mentioned in the main text. Such inhomogeneities can be due to the current density, which is expected to be inhomogeneous since the sample's geometry deviates from a standard Hall bar. Other possible sources could be sample inhomogeneities (disorder or doping for instance) or some thickness variation of the sample. Such a BKT transition was also measured in D4 which is consistant with the ratio $t_1/t_4$ of D1 and D4 thicknesses ($t_1=60$~nm $ > t_4=41$~nm).  We note that any broadening due to heating induced by external radio frequency radiation can be ruled out since a significantly smaller broadening was measured in D4 with the same setup as for D1 as shown in Fig.~3 in the main text.

	\section{Acknowledgements}
	
	FC acknowledges the European Union’s Horizon 2020 research and innovation programme under the Marie Skłodowska-Curie grant agreement No 892728. D.L.B. and Yu.G.N. acknowledge funding by Volkswagen Foundation and are grateful for support by the National Academy of Sciences of Ukraine under project $\Phi 4-19$. J.I.F. acknowledges the support from the Alexander von Humboldt Foundation. S.A. acknowledges financial support by the Deutsche Forschungsgemeinschaft (DFG) through the grant AS 523/4-1. J.D. acknowledges financial support by the Deutsche Forschungsgemeinschaft (DFG) through the SPP 1666 Topological Insulators program (Project DU 1376/2-2) and the Würzburg-Dresden Cluster of Excellence on Complexity and Topology in Quantum Matter - ct.qmat (EXC 2147, project-id 0392019). This project has received funding from the European Research Council (ERC) under the European Unions Horizon 2020 research and innovation program (grant agreement No 647276-MARS-ERC-2014-CoG). We thank Ulrike Nitzsche for technical assistance with the calculations. We would like to thanks Johannes Schoop for his help in sample fabrication.
	

\begin{mcitethebibliography}{46}
	\providecommand*\natexlab[1]{#1}
	\providecommand*\mciteSetBstSublistMode[1]{}
	\providecommand*\mciteSetBstMaxWidthForm[2]{}
	\providecommand*\mciteBstWouldAddEndPuncttrue
	{\def\EndOfBibitem{\unskip.}}
	\providecommand*\mciteBstWouldAddEndPunctfalse
	{\let\EndOfBibitem\relax}
	\providecommand*\mciteSetBstMidEndSepPunct[3]{}
	\providecommand*\mciteSetBstSublistLabelBeginEnd[3]{}
	\providecommand*\EndOfBibitem{}
	\mciteSetBstSublistMode{f}
	\mciteSetBstMaxWidthForm{subitem}{(\alph{mcitesubitemcount})}
	\mciteSetBstSublistLabelBeginEnd
	{\mcitemaxwidthsubitemform\space}
	{\relax}
	{\relax}
	
	\bibitem[Fatemi \latin{et~al.}(2018)Fatemi, Wu, Cao, Bretheau, Gibson,
	Watanabe, Taniguchi, Cava, and Jarillo-Herrero]{Fatemi2018}
	Fatemi,~V.; Wu,~S.; Cao,~Y.; Bretheau,~L.; Gibson,~Q.~D.; Watanabe,~K.;
	Taniguchi,~T.; Cava,~R.~J.; Jarillo-Herrero,~P. Electrically tunable
	low-density superconductivity in a monolayer topological insulator.
	\emph{Science} \textbf{2018}, \emph{362}, 926--929\relax
	\mciteBstWouldAddEndPuncttrue
	\mciteSetBstMidEndSepPunct{\mcitedefaultmidpunct}
	{\mcitedefaultendpunct}{\mcitedefaultseppunct}\relax
	\EndOfBibitem
	\bibitem[Cao \latin{et~al.}(2018)Cao, Fatemi, Fang, Watanabe, Taniguchi,
	Kaxiras, and Jarillo-Herrero]{Cao2018}
	Cao,~Y.; Fatemi,~V.; Fang,~S.; Watanabe,~K.; Taniguchi,~T.; Kaxiras,~E.;
	Jarillo-Herrero,~P. Unconventional superconductivity in magic-angle graphene
	superlattices. \emph{Nature} \textbf{2018}, \emph{556}, 43\relax
	\mciteBstWouldAddEndPuncttrue
	\mciteSetBstMidEndSepPunct{\mcitedefaultmidpunct}
	{\mcitedefaultendpunct}{\mcitedefaultseppunct}\relax
	\EndOfBibitem
	\bibitem[Li \latin{et~al.}(2016)Li, Wang, Yao, and Lee]{Li2016c}
	Li,~Z.-X.; Wang,~F.; Yao,~H.; Lee,~D.-H. What makes the $T_\mathrm{c}$ of
	monolayer FeSe on SrTiO$_3$ so high: a sign-problem-free quantum Monte Carlo
	study. \emph{Science Bulletin} \textbf{2016}, \emph{61}, 925--930\relax
	\mciteBstWouldAddEndPuncttrue
	\mciteSetBstMidEndSepPunct{\mcitedefaultmidpunct}
	{\mcitedefaultendpunct}{\mcitedefaultseppunct}\relax
	\EndOfBibitem
	\bibitem[Rhodes \latin{et~al.}(2021)Rhodes, Jindal, Yuan, Jung, Antony, Wang,
	Kim, Chiu, Taniguchi, Watanabe, Barmak, Balicas, Dean, Qian, Fu, Pasupathy,
	and Hone]{Rhodes2021}
	Rhodes,~D.~A. \latin{et~al.}  Enhanced Superconductivity in Monolayer
	Td-MoTe$_2$. \emph{Nano Lett.} \textbf{2021}, \emph{21}, 2505--2511\relax
	\mciteBstWouldAddEndPuncttrue
	\mciteSetBstMidEndSepPunct{\mcitedefaultmidpunct}
	{\mcitedefaultendpunct}{\mcitedefaultseppunct}\relax
	\EndOfBibitem
	\bibitem[Cao \latin{et~al.}(2021)Cao, Park, Watanabe, Taniguchi, and
	Jarillo-Herrero]{Cao2021}
	Cao,~Y.; Park,~J.~M.; Watanabe,~K.; Taniguchi,~T.; Jarillo-Herrero,~P.
	Pauli-limit violation and re-entrant superconductivity in moiré graphene.
	\emph{Nature} \textbf{2021}, \emph{595}, 526--531\relax
	\mciteBstWouldAddEndPuncttrue
	\mciteSetBstMidEndSepPunct{\mcitedefaultmidpunct}
	{\mcitedefaultendpunct}{\mcitedefaultseppunct}\relax
	\EndOfBibitem
	\bibitem[Berezinskii(1972)]{Berezinskii1972}
	Berezinskii,~V.~L. Destruction of Long-range Order in One-dimensional and
	Two-dimensional Systems Possessing a Continuous Symmetry Group. II. Quantum
	Systems. \emph{Soviet Journal of Experimental and Theoretical Physics}
	\textbf{1972}, \emph{34}, 610\relax
	\mciteBstWouldAddEndPuncttrue
	\mciteSetBstMidEndSepPunct{\mcitedefaultmidpunct}
	{\mcitedefaultendpunct}{\mcitedefaultseppunct}\relax
	\EndOfBibitem
	\bibitem[Kosterlitz and Thouless(1973)Kosterlitz, and Thouless]{Kosterlitz1973}
	Kosterlitz,~J.~M.; Thouless,~D.~J. Ordering, metastability and phase
	transitions in two-dimensional systems. \emph{Journal of Physics C: Solid
		State Physics} \textbf{1973}, \emph{6}, 1181--1203\relax
	\mciteBstWouldAddEndPuncttrue
	\mciteSetBstMidEndSepPunct{\mcitedefaultmidpunct}
	{\mcitedefaultendpunct}{\mcitedefaultseppunct}\relax
	\EndOfBibitem
	\bibitem[Edel'shtein(1989)]{Edelshtein1989}
	Edel'shtein,~V.~M. Characteristics of the Cooper pairing in two-dimensional
	noncentrosymmetric electron systems. \emph{Soviet Physics - JETP (English
		Translation)} \textbf{1989}, \emph{68}, 1244--1249\relax
	\mciteBstWouldAddEndPuncttrue
	\mciteSetBstMidEndSepPunct{\mcitedefaultmidpunct}
	{\mcitedefaultendpunct}{\mcitedefaultseppunct}\relax
	\EndOfBibitem
	\bibitem[Gor'kov and Rashba(2001)Gor'kov, and Rashba]{Gorkov2001}
	Gor'kov,~L.~P.; Rashba,~E.~I. Superconducting 2D System with Lifted Spin
	Degeneracy: Mixed Singlet-Triplet State. \emph{Phys. Rev. Lett.}
	\textbf{2001}, \emph{87}, 037004\relax
	\mciteBstWouldAddEndPuncttrue
	\mciteSetBstMidEndSepPunct{\mcitedefaultmidpunct}
	{\mcitedefaultendpunct}{\mcitedefaultseppunct}\relax
	\EndOfBibitem
	\bibitem[Kozii and Fu(2015)Kozii, and Fu]{Kozii2015}
	Kozii,~V.; Fu,~L. Odd-Parity Superconductivity in the Vicinity of Inversion
	Symmetry Breaking in Spin-Orbit-Coupled Systems. \emph{Phys. Rev. Lett.}
	\textbf{2015}, \emph{115}, 207002\relax
	\mciteBstWouldAddEndPuncttrue
	\mciteSetBstMidEndSepPunct{\mcitedefaultmidpunct}
	{\mcitedefaultendpunct}{\mcitedefaultseppunct}\relax
	\EndOfBibitem
	\bibitem[Fulde and Ferrell(1964)Fulde, and Ferrell]{Fulde1964}
	Fulde,~P.; Ferrell,~R.~A. Superconductivity in a Strong Spin-Exchange Field.
	\emph{Phys. Rev.} \textbf{1964}, \emph{135}, A550--A563\relax
	\mciteBstWouldAddEndPuncttrue
	\mciteSetBstMidEndSepPunct{\mcitedefaultmidpunct}
	{\mcitedefaultendpunct}{\mcitedefaultseppunct}\relax
	\EndOfBibitem
	\bibitem[Larkin and Ovchinnikov(1965)Larkin, and Ovchinnikov]{Larkin1965}
	Larkin,~A.~I.; Ovchinnikov,~I. U.~N. Inhomogeneous state of
	superconductors(Production of superconducting state in ferromagnet with Fermi
	surfaces, examining Green function). \emph{Soviet Physics-JETP}
	\textbf{1965}, \emph{20}, 762--769\relax
	\mciteBstWouldAddEndPuncttrue
	\mciteSetBstMidEndSepPunct{\mcitedefaultmidpunct}
	{\mcitedefaultendpunct}{\mcitedefaultseppunct}\relax
	\EndOfBibitem
	\bibitem[Mayaffre \latin{et~al.}(2014)Mayaffre, Krämer, Horvatić, Berthier,
	Miyagawa, Kanoda, and Mitrović]{Mayaffre2014}
	Mayaffre,~H.; Krämer,~S.; Horvatić,~M.; Berthier,~C.; Miyagawa,~K.;
	Kanoda,~K.; Mitrović,~V.~F. Evidence of Andreev bound states as a hallmark
	of the FFLO phase in $\kappa$-(BEDT-TTF)$_2$Cu(NCS)$_2$. \emph{Nature
		Physics} \textbf{2014}, \emph{10}, 928\relax
	\mciteBstWouldAddEndPuncttrue
	\mciteSetBstMidEndSepPunct{\mcitedefaultmidpunct}
	{\mcitedefaultendpunct}{\mcitedefaultseppunct}\relax
	\EndOfBibitem
	\bibitem[Hosur \latin{et~al.}(2014)Hosur, Dai, Fang, and Qi]{Hosur2014}
	Hosur,~P.; Dai,~X.; Fang,~Z.; Qi,~X.-L. Time-reversal-invariant topological
	superconductivity in doped Weyl semimetals. \emph{Phys. Rev. B}
	\textbf{2014}, \emph{90}, 045130\relax
	\mciteBstWouldAddEndPuncttrue
	\mciteSetBstMidEndSepPunct{\mcitedefaultmidpunct}
	{\mcitedefaultendpunct}{\mcitedefaultseppunct}\relax
	\EndOfBibitem
	\bibitem[Bednik \latin{et~al.}(2015)Bednik, Zyuzin, and Burkov]{Bednik2015}
	Bednik,~G.; Zyuzin,~A.~A.; Burkov,~A.~A. Superconductivity in Weyl metals.
	\emph{Phys. Rev. B} \textbf{2015}, \emph{92}, 035153\relax
	\mciteBstWouldAddEndPuncttrue
	\mciteSetBstMidEndSepPunct{\mcitedefaultmidpunct}
	{\mcitedefaultendpunct}{\mcitedefaultseppunct}\relax
	\EndOfBibitem
	\bibitem[Wang \latin{et~al.}(2018)Wang, Kong, Fan, Chen, Zhu, Liu, Cao, Sun,
	Du, Schneeloch, Zhong, Gu, Fu, Ding, and Gao]{Wang2018d}
	Wang,~D.; Kong,~L.; Fan,~P.; Chen,~H.; Zhu,~S.; Liu,~W.; Cao,~L.; Sun,~Y.;
	Du,~S.; Schneeloch,~J.; Zhong,~R.; Gu,~G.; Fu,~L.; Ding,~H.; Gao,~H.-J.
	Evidence for Majorana bound states in an iron-based superconductor.
	\emph{Science} \textbf{2018}, \emph{362}, 333--335\relax
	\mciteBstWouldAddEndPuncttrue
	\mciteSetBstMidEndSepPunct{\mcitedefaultmidpunct}
	{\mcitedefaultendpunct}{\mcitedefaultseppunct}\relax
	\EndOfBibitem
	\bibitem[Tang \latin{et~al.}(2019)Tang, Wang, Wang, Gan, Gu, Zhang, He, and
	Zhang]{Tang2019}
	Tang,~F.; Wang,~P.; Wang,~P.; Gan,~Y.; Gu,~G.~D.; Zhang,~W.; He,~M.; Zhang,~L.
	Quasi-2D superconductivity in FeTe$_{0.55}$Se$_{0.45}$ ultrathin film.
	\emph{Journal of Physics: Condensed Matter} \textbf{2019}, \emph{31},
	265702\relax
	\mciteBstWouldAddEndPuncttrue
	\mciteSetBstMidEndSepPunct{\mcitedefaultmidpunct}
	{\mcitedefaultendpunct}{\mcitedefaultseppunct}\relax
	\EndOfBibitem
	\bibitem[Brun \latin{et~al.}(2014)Brun, Cren, Cherkez, Debontridder, Pons,
	Fokin, Tringides, Bozhko, Ioffe, Altshuler, and Roditchev]{Brun2014}
	Brun,~C.; Cren,~T.; Cherkez,~V.; Debontridder,~F.; Pons,~S.; Fokin,~D.;
	Tringides,~M.~C.; Bozhko,~S.; Ioffe,~L.~B.; Altshuler,~B.~L.; Roditchev,~D.
	Remarkable effects of disorder on superconductivity of single atomic layers
	of lead on silicon. \emph{Nature Physics} \textbf{2014}, \emph{10},
	444--450\relax
	\mciteBstWouldAddEndPuncttrue
	\mciteSetBstMidEndSepPunct{\mcitedefaultmidpunct}
	{\mcitedefaultendpunct}{\mcitedefaultseppunct}\relax
	\EndOfBibitem
	\bibitem[Brun \latin{et~al.}(2016)Brun, Cren, and Roditchev]{Brun2016}
	Brun,~C.; Cren,~T.; Roditchev,~D. Review of 2D superconductivity: the ultimate
	case of epitaxial monolayers. \emph{Superconductor Science and Technology}
	\textbf{2016}, \emph{30}, 013003\relax
	\mciteBstWouldAddEndPuncttrue
	\mciteSetBstMidEndSepPunct{\mcitedefaultmidpunct}
	{\mcitedefaultendpunct}{\mcitedefaultseppunct}\relax
	\EndOfBibitem
	\bibitem[Yang \latin{et~al.}(2016)Yang, Bai, Wang, Li, Chen, Chen, Li, Feng,
	Zheng, and Xu]{Yang2016}
	Yang,~X.; Bai,~H.; Wang,~Z.; Li,~Y.; Chen,~Q.; Chen,~J.; Li,~Y.; Feng,~C.;
	Zheng,~Y.; Xu,~Z.-a. Giant linear magneto-resistance in nonmagnetic PtBi$_2$.
	\emph{Appl. Phys. Lett.} \textbf{2016}, \emph{108}, 252401\relax
	\mciteBstWouldAddEndPuncttrue
	\mciteSetBstMidEndSepPunct{\mcitedefaultmidpunct}
	{\mcitedefaultendpunct}{\mcitedefaultseppunct}\relax
	\EndOfBibitem
	\bibitem[Gao \latin{et~al.}(2017)Gao, Hao, Zheng, Ning, Wu, Zhu, Zheng, Zhang,
	Lu, Zhang, Xi, Yang, Du, Zhang, Zhang, and Tian]{Gao2017}
	Gao,~W. \latin{et~al.}  Extremely Large Magnetoresistance in a Topological
	Semimetal Candidate Pyrite ${\mathrm{PtBi}}_{2}$. \emph{Phys. Rev. Lett.}
	\textbf{2017}, \emph{118}, 256601\relax
	\mciteBstWouldAddEndPuncttrue
	\mciteSetBstMidEndSepPunct{\mcitedefaultmidpunct}
	{\mcitedefaultendpunct}{\mcitedefaultseppunct}\relax
	\EndOfBibitem
	\bibitem[Feng \latin{et~al.}(2019)Feng, Jiang, Feng, Yang, Xu, Liu, Yang,
	Arita, Schwier, Shimada, Jeschke, Thomale, Shi, Wu, Xiao, Qiao, and
	He]{Feng2019}
	Feng,~Y. \latin{et~al.}  Rashba-like spin splitting along three momentum
	directions in trigonal layered PtBi$_2$. \emph{Nature Communications}
	\textbf{2019}, \emph{10}, 4765\relax
	\mciteBstWouldAddEndPuncttrue
	\mciteSetBstMidEndSepPunct{\mcitedefaultmidpunct}
	{\mcitedefaultendpunct}{\mcitedefaultseppunct}\relax
	\EndOfBibitem
	\bibitem[Gao \latin{et~al.}(2018)Gao, Zhu, Zheng, Wu, Zhang, Xi, Zhang, Zhang,
	Hao, Ning, and Tian]{Gao2018}
	Gao,~W.; Zhu,~X.; Zheng,~F.; Wu,~M.; Zhang,~J.; Xi,~C.; Zhang,~P.; Zhang,~Y.;
	Hao,~N.; Ning,~W.; Tian,~M. A possible candidate for triply degenerate point
	fermions in trigonal layered PtBi$_2$. \emph{Nature Communications}
	\textbf{2018}, \emph{9}, 3249\relax
	\mciteBstWouldAddEndPuncttrue
	\mciteSetBstMidEndSepPunct{\mcitedefaultmidpunct}
	{\mcitedefaultendpunct}{\mcitedefaultseppunct}\relax
	\EndOfBibitem
	\bibitem[Nie \latin{et~al.}(2020)Nie, Li, Yang, Zhu, Xu, Yang, Zheng, Guan,
	Wang, Li, Liu, Li, Zhang, Shi, Zheng, and Jia]{Nie2020}
	Nie,~X.-A. \latin{et~al.}  Robust Hot Electron and Multiple Topological
	Insulator States in PtBi$_2$. \emph{ACS Nano} \textbf{2020}, \emph{14},
	2366--2372\relax
	\mciteBstWouldAddEndPuncttrue
	\mciteSetBstMidEndSepPunct{\mcitedefaultmidpunct}
	{\mcitedefaultendpunct}{\mcitedefaultseppunct}\relax
	\EndOfBibitem
	\bibitem[Wang \latin{et~al.}(2021)Wang, Chen, Zhou, An, Zhou, Gu, Tian, and
	Yang]{Wang2021}
	Wang,~J.; Chen,~X.; Zhou,~Y.; An,~C.; Zhou,~Y.; Gu,~C.; Tian,~M.; Yang,~Z.
	Pressure-induced superconductivity in trigonal layered ${\mathrm{PtBi}}_{2}$
	with triply degenerate point fermions. \emph{Phys. Rev. B} \textbf{2021},
	\emph{103}, 014507\relax
	\mciteBstWouldAddEndPuncttrue
	\mciteSetBstMidEndSepPunct{\mcitedefaultmidpunct}
	{\mcitedefaultendpunct}{\mcitedefaultseppunct}\relax
	\EndOfBibitem
	\bibitem[Bashlakov \latin{et~al.}(2022)Bashlakov, Kvitnitskaya, Shipunov,
	Aswartham, Feya, Efremov, Büchner, and Naidyuk]{Bashlakov2022}
	Bashlakov,~D.~L.; Kvitnitskaya,~O.~E.; Shipunov,~G.; Aswartham,~S.;
	Feya,~O.~D.; Efremov,~D.~V.; Büchner,~B.; Naidyuk,~Y.~G. Electron-phonon
	interaction and point contact enhanced superconductivity in trigonal
	PtBi$_2$. \emph{Low Temperature Physics} \textbf{2022}, \emph{48},
	747--754\relax
	\mciteBstWouldAddEndPuncttrue
	\mciteSetBstMidEndSepPunct{\mcitedefaultmidpunct}
	{\mcitedefaultendpunct}{\mcitedefaultseppunct}\relax
	\EndOfBibitem
	\bibitem[Shipunov \latin{et~al.}(2020)Shipunov, Kovalchuk, Piening,
	Labracherie, Veyrat, Wolf, Lubk, Subakti, Giraud, Dufouleur, Shokri,
	Caglieris, Hess, Efremov, B\"uchner, and Aswartham]{Shipunov2020}
	Shipunov,~G. \latin{et~al.}  Polymorphic ${\mathrm{PtBi}}_{2}$: Growth,
	structure, and superconducting properties. \emph{Phys. Rev. Materials}
	\textbf{2020}, \emph{4}, 124202\relax
	\mciteBstWouldAddEndPuncttrue
	\mciteSetBstMidEndSepPunct{\mcitedefaultmidpunct}
	{\mcitedefaultendpunct}{\mcitedefaultseppunct}\relax
	\EndOfBibitem
	\bibitem[Koepernik and Eschrig(1999)Koepernik, and Eschrig]{Koepernik1999}
	Koepernik,~K.; Eschrig,~H. Full-potential nonorthogonal local-orbital
	minimum-basis band-structure scheme. \emph{Phys. Rev. B} \textbf{1999},
	\emph{59}, 1743--1757\relax
	\mciteBstWouldAddEndPuncttrue
	\mciteSetBstMidEndSepPunct{\mcitedefaultmidpunct}
	{\mcitedefaultendpunct}{\mcitedefaultseppunct}\relax
	\EndOfBibitem
	\bibitem[Tinkham(1975)]{Tinkham1975}
	Tinkham,~M. \emph{Introduction to superconductivity}; International series in
	pure and applied physics; McGraw-Hill: New York, 1975\relax
	\mciteBstWouldAddEndPuncttrue
	\mciteSetBstMidEndSepPunct{\mcitedefaultmidpunct}
	{\mcitedefaultendpunct}{\mcitedefaultseppunct}\relax
	\EndOfBibitem
	\bibitem[Clogston(1962)]{Clogston1962}
	Clogston,~A.~M. Upper Limit for the Critical Field in Hard Superconductors.
	\emph{Phys. Rev. Lett.} \textbf{1962}, \emph{9}, 266--267\relax
	\mciteBstWouldAddEndPuncttrue
	\mciteSetBstMidEndSepPunct{\mcitedefaultmidpunct}
	{\mcitedefaultendpunct}{\mcitedefaultseppunct}\relax
	\EndOfBibitem
	\bibitem[Chandrasekhar(1962)]{Chandrasekhar1962}
	Chandrasekhar,~B.~S. A note on the maximum critical field of high field
	superconductors. \emph{Appl. Phys. Lett.} \textbf{1962}, \emph{1}, 7--8\relax
	\mciteBstWouldAddEndPuncttrue
	\mciteSetBstMidEndSepPunct{\mcitedefaultmidpunct}
	{\mcitedefaultendpunct}{\mcitedefaultseppunct}\relax
	\EndOfBibitem
	\bibitem[Cui \latin{et~al.}(2019)Cui, Li, Zhou, He, Huang, Yi, Fan, Ji, Jing,
	Qu, Cheng, Yang, Lu, Suenaga, Liu, Law, Lin, Liu, and Liu]{Cui2019}
	Cui,~J. \latin{et~al.}  Transport evidence of asymmetric spin-orbit coupling in
	few-layer superconducting 1Td-MoTe$_2$. \emph{Nature Communications}
	\textbf{2019}, \emph{10}, 2044\relax
	\mciteBstWouldAddEndPuncttrue
	\mciteSetBstMidEndSepPunct{\mcitedefaultmidpunct}
	{\mcitedefaultendpunct}{\mcitedefaultseppunct}\relax
	\EndOfBibitem
	\bibitem[Benfatto \latin{et~al.}(2009)Benfatto, Castellani, and
	Giamarchi]{Benfatto2009}
	Benfatto,~L.; Castellani,~C.; Giamarchi,~T. Broadening of the
	Berezinskii-Kosterlitz-Thouless superconducting transition by inhomogeneity
	and finite-size effects. \emph{Phys. Rev. B} \textbf{2009}, \emph{80},
	214506\relax
	\mciteBstWouldAddEndPuncttrue
	\mciteSetBstMidEndSepPunct{\mcitedefaultmidpunct}
	{\mcitedefaultendpunct}{\mcitedefaultseppunct}\relax
	\EndOfBibitem
	\bibitem[Halperin and Nelson(1979)Halperin, and Nelson]{Halperin1979}
	Halperin,~B.~I.; Nelson,~D.~R. Resistive transition in superconducting films.
	\emph{Journal of Low Temperature Physics} \textbf{1979}, \emph{36},
	599--616\relax
	\mciteBstWouldAddEndPuncttrue
	\mciteSetBstMidEndSepPunct{\mcitedefaultmidpunct}
	{\mcitedefaultendpunct}{\mcitedefaultseppunct}\relax
	\EndOfBibitem
	\bibitem[Baity \latin{et~al.}(2016)Baity, Shi, Shi, Benfatto, and
	Popovi\ifmmode~\acute{c}\else \'{c}\fi{}]{Baity2016}
	Baity,~P.~G.; Shi,~X.; Shi,~Z.; Benfatto,~L.; Popovi\ifmmode~\acute{c}\else
	\'{c}\fi{},~D. Effective two-dimensional thickness for the
	Berezinskii-Kosterlitz-Thouless-like transition in a highly underdoped
	${\mathrm{La}}_{2\ensuremath{-}x}{\mathrm{Sr}}_{x}{\mathrm{CuO}}_{4}$.
	\emph{Phys. Rev. B} \textbf{2016}, \emph{93}, 024519\relax
	\mciteBstWouldAddEndPuncttrue
	\mciteSetBstMidEndSepPunct{\mcitedefaultmidpunct}
	{\mcitedefaultendpunct}{\mcitedefaultseppunct}\relax
	\EndOfBibitem
	\bibitem[Guo \latin{et~al.}(2017)Guo, Chen, Jia, Zhang, Liu, Lei, Li, Gu, Jin,
	and Chen]{Guo2017}
	Guo,~J.~G.; Chen,~X.; Jia,~X.~Y.; Zhang,~Q.~H.; Liu,~N.; Lei,~H.~C.; Li,~S.~Y.;
	Gu,~L.; Jin,~S.~F.; Chen,~X.~L. Quasi-two-dimensional superconductivity from
	dimerization of atomically ordered AuTe(2)Se(4/3) cubes. \emph{Nature
		communications} \textbf{2017}, \emph{8}, 871--871\relax
	\mciteBstWouldAddEndPuncttrue
	\mciteSetBstMidEndSepPunct{\mcitedefaultmidpunct}
	{\mcitedefaultendpunct}{\mcitedefaultseppunct}\relax
	\EndOfBibitem
	\bibitem[Mondal \latin{et~al.}(2011)Mondal, Kumar, Chand, Kamlapure, Saraswat,
	Seibold, Benfatto, and Raychaudhuri]{Mondal2011}
	Mondal,~M.; Kumar,~S.; Chand,~M.; Kamlapure,~A.; Saraswat,~G.; Seibold,~G.;
	Benfatto,~L.; Raychaudhuri,~P. Role of the Vortex-Core Energy on the
	Berezinskii-Kosterlitz-Thouless Transition in Thin Films of NbN. \emph{Phys.
		Rev. Lett.} \textbf{2011}, \emph{107}, 217003\relax
	\mciteBstWouldAddEndPuncttrue
	\mciteSetBstMidEndSepPunct{\mcitedefaultmidpunct}
	{\mcitedefaultendpunct}{\mcitedefaultseppunct}\relax
	\EndOfBibitem
	\bibitem[Kaiser \latin{et~al.}(2014)Kaiser, Baranov, and Ruck]{Kaiser2014}
	Kaiser,~M.; Baranov,~A.~I.; Ruck,~M. Bi2Pt(hP9) by Low-Temperature Reduction of
	Bi13Pt3I7: Reinvestigation of the Crystal Structure and Chemical Bonding
	Analysis. \emph{Z. anorg. allg. Chem.} \textbf{2014}, \emph{640},
	2742--2746\relax
	\mciteBstWouldAddEndPuncttrue
	\mciteSetBstMidEndSepPunct{\mcitedefaultmidpunct}
	{\mcitedefaultendpunct}{\mcitedefaultseppunct}\relax
	\EndOfBibitem
	\bibitem[Xing \latin{et~al.}(2020)Xing, Chapai, Nepal, and Jin]{Xing2020}
	Xing,~L.; Chapai,~R.; Nepal,~R.; Jin,~R. Topological behavior and Zeeman
	splitting in trigonal PtBi2-x single crystals. \emph{npj Quantum Materials}
	\textbf{2020}, \emph{5}, 10\relax
	\mciteBstWouldAddEndPuncttrue
	\mciteSetBstMidEndSepPunct{\mcitedefaultmidpunct}
	{\mcitedefaultendpunct}{\mcitedefaultseppunct}\relax
	\EndOfBibitem
	\bibitem[de~Gennes(1966)]{Gennes1966}
	de~Gennes,~P.~G. \emph{Superconductivity of Metals and Alloys}; Benjamin: New
	York, 1966\relax
	\mciteBstWouldAddEndPuncttrue
	\mciteSetBstMidEndSepPunct{\mcitedefaultmidpunct}
	{\mcitedefaultendpunct}{\mcitedefaultseppunct}\relax
	\EndOfBibitem
	\bibitem[Hu(1972)]{Hu1972}
	Hu,~C.-R. Numerical Constants for Isolated Vortices in Superconductors.
	\emph{Phys. Rev. B} \textbf{1972}, \emph{6}, 1756--1760\relax
	\mciteBstWouldAddEndPuncttrue
	\mciteSetBstMidEndSepPunct{\mcitedefaultmidpunct}
	{\mcitedefaultendpunct}{\mcitedefaultseppunct}\relax
	\EndOfBibitem
	\bibitem[Alama \latin{et~al.}(1999)Alama, Berlinsky, Bronsard, and
	Giorgi]{Alama1999}
	Alama,~S.; Berlinsky,~A.~J.; Bronsard,~L.; Giorgi,~T. Vortices with
	antiferromagnetic cores in the SO(5) model of high-temperature
	superconductivity. \emph{Phys. Rev. B} \textbf{1999}, \emph{60},
	6901--6906\relax
	\mciteBstWouldAddEndPuncttrue
	\mciteSetBstMidEndSepPunct{\mcitedefaultmidpunct}
	{\mcitedefaultendpunct}{\mcitedefaultseppunct}\relax
	\EndOfBibitem
	\bibitem[Benfatto \latin{et~al.}(2007)Benfatto, Castellani, and
	Giamarchi]{Benfatto2007}
	Benfatto,~L.; Castellani,~C.; Giamarchi,~T. Kosterlitz-Thouless Behavior in
	Layered Superconductors: The Role of the Vortex Core Energy. \emph{Phys. Rev.
		Lett.} \textbf{2007}, \emph{98}, 117008\relax
	\mciteBstWouldAddEndPuncttrue
	\mciteSetBstMidEndSepPunct{\mcitedefaultmidpunct}
	{\mcitedefaultendpunct}{\mcitedefaultseppunct}\relax
	\EndOfBibitem
	\bibitem[Benfatto \latin{et~al.}(2008)Benfatto, Castellani, and
	Giamarchi]{Benfatto2008}
	Benfatto,~L.; Castellani,~C.; Giamarchi,~T. Doping dependence of the
	vortex-core energy in bilayer films of cuprates. \emph{Phys. Rev. B}
	\textbf{2008}, \emph{77}, 100506\relax
	\mciteBstWouldAddEndPuncttrue
	\mciteSetBstMidEndSepPunct{\mcitedefaultmidpunct}
	{\mcitedefaultendpunct}{\mcitedefaultseppunct}\relax
	\EndOfBibitem
	\bibitem[Vaglio \latin{et~al.}(1993)Vaglio, Attanasio, Maritato, and
	Ruosi]{Vaglio1993}
	Vaglio,~R.; Attanasio,~C.; Maritato,~L.; Ruosi,~A. Explanation of the
	resistance-peak anomaly in nonhomogeneous superconductors. \emph{Phys. Rev.
		B} \textbf{1993}, \emph{47}, 15302--15303\relax
	\mciteBstWouldAddEndPuncttrue
	\mciteSetBstMidEndSepPunct{\mcitedefaultmidpunct}
	{\mcitedefaultendpunct}{\mcitedefaultseppunct}\relax
	\EndOfBibitem
\end{mcitethebibliography}

\providecommand{\latin}[1]{#1}
\makeatletter
\providecommand{\doi}
{\begingroup\let\do\@makeother\dospecials
	\catcode`\{=1 \catcode`\}=2 \doi@aux}
\providecommand{\doi@aux}[1]{\endgroup\texttt{#1}}
\makeatother
\providecommand*\mcitethebibliography{\thebibliography}
\csname @ifundefined\endcsname{endmcitethebibliography}
{\let\endmcitethebibliography\endthebibliography}{}

\end{document}